\documentclass[aps, prx, reprint, superscriptaddress, floatfix]{revtex4-2}
\pdfoutput=1
\usepackage{braket,dsfont,subfigure,amsfonts,amssymb,bm,graphicx,amsmath,amsthm,txfonts}
\usepackage[dvipsnames]{xcolor}
\usepackage[colorlinks]{hyperref}
\usepackage{enumitem}
\usepackage{multirow}
\usepackage{orcidlink}
\newcommand{\Tr}{{\rm Tr}}

\begin{document}

\title{Tensor network characterization and mitigation of readout errors}
\author{Yuchen Guo~\orcidlink{0000-0002-4901-2737}}
\affiliation{State Key Laboratory of Low Dimensional Quantum Physics and Department of Physics, Tsinghua University, Beijing 100084, China}
\author{Shuo Yang~\orcidlink{0000-0001-9733-8566}}
\email{shuoyang@tsinghua.edu.cn}
\affiliation{State Key Laboratory of Low Dimensional Quantum Physics and Department of Physics, Tsinghua University, Beijing 100084, China}
\affiliation{Frontier Science Center for Quantum Information, Beijing 100084, China}
\affiliation{Hefei National Laboratory, Hefei 230088, China}

\begin{abstract}
Readout errors are a major bottleneck to extracting reliable information from near-term quantum processors, especially when spatial correlations are non-negligible.
We present a unified tensor-network framework that models the readout process as a matrix product operator (MPO), enabling efficient characterization and mitigation beyond uncorrelated approximations.
The MPO model is trained via likelihood optimization on calibration data and applies to multiple tasks, including nonlocal observable estimation, random circuit sampling, and random-measurement protocols, such as classical shadows and learning-based tomography.
Experiments on a superconducting processor and numerical simulations up to 20 qubits show that the MPO model captures correlated readout errors that uncorrelated models miss, with a sample cost that grows only near-linearly with system size.
When extended to two-dimensional systems, the framework can also be integrated with tensor-network quantum error-correction decoders by performing joint inference over data and readout errors.
These results establish tensor-network readout error mitigation as a scalable and versatile approach for noise-aware quantum data processing.

\end{abstract}

\maketitle

\tableofcontents

\section{Introduction}
Quantum processors have witnessed rapid improvements in qubit number and gate fidelity, enabling increasingly complex experiments in quantum simulation, variational algorithms, and random circuit sampling~\cite{Kandala2017,Arute2019,Cerezo2021}.
However, readout errors remain one of the dominant noise sources on current noisy intermediate-scale quantum (NISQ) devices~\cite{Preskill2018,Maciejewski2020,Bravyi2021}.
Unlike coherent gate errors, readout errors directly distort the final classical outcomes, and therefore bias almost all tasks that rely on measurement data, including expectation-value estimation, tomography, and sampling-based benchmarks~\cite{Huang2020,Torlai2023}.

A standard approach to readout error mitigation (REM) is to characterize a classical transition matrix $\mathbf{\Lambda}$ that maps ideal measurement distributions to noisy ones~\cite{Maciejewski2020,Chen2019}.
Full characterization of $\mathbf{\Lambda}$ requires resources exponential in system size.
To avoid this exponential cost, widely used factorized models assume independent single-qubit readout processes~\cite{Maciejewski2020}.
Such uncorrelated models neglect spatial correlations induced by crosstalk, shared control lines, and hardware-dependent measurement mechanisms~\cite{Maciejewski2021,Bravyi2021}.
These correlated effects are often non-negligible in real devices and can limit mitigation performance, especially for nonlocal observables and large-scale sampling tasks.

In this work, we develop a unified tensor-network framework for characterization and mitigation of correlated readout errors.
Our key idea is to represent the readout error matrix by a matrix product operator (MPO), which naturally captures local and short-range correlations with polynomial complexity~\cite{Schollwoeck2011,Orus2014}.
We propose a likelihood-based learning method to train the MPO directly from calibration data, together with an efficient tensor-contraction algorithm for optimization.
Built on this learned model, we show how to perform REM in multiple practically relevant settings: (i) mitigation of nonlocal observables via efficient contraction with an MPO approximation of $\mathbf{\Lambda}^{-1}$, (ii) mitigation of global sampling tasks by learning an ideal-distribution model under the noisy likelihood, (iii) mitigation under random measurements, including classical-shadow estimation and learning-based quantum state tomography, and (iv) integration with tensor-network quantum error correction (QEC) decoding, where the learned readout error matrix is combined with the decoder's likelihood model to directly infer the most probable logical class.

We validate the framework through experiments on a superconducting quantum processor and extensive numerical simulations up to 20 qubits.
Experimentally, the MPO model consistently outperforms the uncorrelated model under the same measurement cost, demonstrating clear evidence of correlated readout noise in real devices.
Numerically, the required sample number scales nearly linearly with system size, in sharp contrast to the exponential cost of full characterization, and the mitigation performance remains robust across different tasks.
These results indicate that tensor-network REM provides a scalable route toward more reliable extraction of physical information from both near-term and fault-tolerant quantum devices.

\section{Characterization and mitigation of readout errors}
Suppose an $N$-qubit quantum system is measured in the computational basis $\{\ket{0}, \ket{1}\}^{\otimes N}$.
The readout error can be described by a $2^N \times 2^N$ matrix $\mathbf{\Lambda}$, whose matrix element $\mathbf{\Lambda}_{\mathbf{x},\mathbf{y}}$ represents the probability of obtaining the measurement outcome $\mathbf{x}=\{x_1, x_2, \cdots, x_N\}$ when the actual result is $\mathbf{y}$. 
Therefore, the relationship between the ideal probability distribution $\mathbf{P}_{\mathrm{ideal}}$ and the noisy one $\mathbf{P}_{\mathrm{noisy}}$ (both in the computational basis) can be expressed as~\cite{Maciejewski2020}
\begin{align}
    \mathbf{P}_{\mathrm{noisy}} = \mathbf{\Lambda} \mathbf{P}_{\mathrm{ideal}}.
\end{align}

\subsection{Characterization of the readout error model}
The first step is to characterize the readout error matrix $\mathbf{\Lambda}$.
In general, $\mathbf{\Lambda}$ can be obtained by preparing and measuring all $2^N$ computational basis states (a process also known as quantum detector tomography~\cite{Chen2019}), which is infeasible for large $N$ due to the exponential scaling.
Most works assume that the readout errors are independent for each qubit, leading to a tensor product structure of $\mathbf{\Lambda}$, i.e.,~\cite{Maciejewski2020}
\begin{align}
    \mathbf{\Lambda} = \bigotimes_{k=1}^N \mathbf{\Lambda}^{[k]},
\end{align}
where $\mathbf{\Lambda}^{[k]}$ is the $2 \times 2$ readout error matrix for the $k$-th qubit.
In this case, $\mathbf{\Lambda}$ can be efficiently characterized by preparing $\ket{0}$ and $\ket{1}$ states and collecting the measurement outcomes individually for each qubit.

However, this assumption neglects the correlated readout errors among different qubits, which can be significant in practical quantum devices, especially when qubits are closely spaced and crosstalk errors occur.
Some works extend this structure by further considering the correlation between different qubits by grouping the qubits into several clusters, where $\mathbf{\Lambda}$ is expressed as a tensor product of the readout error matrices of these clusters~\cite{Maciejewski2021}.
A correlated Markovian model has also been proposed to capture the nearest-neighbor correlations among qubits arranged in a one-dimensional (1D) chain~\cite{Bravyi2021}.
As discussed in the following, different applications impose various requirements and constraints on the structure of the error model, necessitating specific adaptations and designs.
Therefore, a unified framework is desired to efficiently characterize the readout error matrix $\mathbf{\Lambda}$ with local and short-range correlations among qubits, such that they can be naturally combined into the following processes of REM in various applications.

\subsection{Mitigation of readout errors}
\subsubsection{Local distribution and observables}
After the readout error matrix $\mathbf{\Lambda}$ is characterized, the next step is to mitigate the readout errors in real experiments.
For those few-qubit measurements, a complete distribution $\mathbf{P}_{\mathrm{noisy}}$ can be obtained by repeatedly preparing and measuring the quantum state.
Then, the ideal distribution $\mathbf{P}_{\mathrm{ideal}}$ can be estimated by inverting the readout error matrix $\mathbf{\Lambda}$ as
\begin{align}
    \mathbf{P}_{\mathrm{ideal}} = \mathbf{\Lambda}^{-1} \mathbf{P}_{\mathrm{noisy}}.\label{Equ: Inverse}
\end{align}
To avoid the instability of matrix inversion, various classical post-processing methods have also been proposed, such as least squares fitting~\cite{Chen2019}
\begin{align}
    \mathbf{P}_{\mathrm{ideal}} = \arg \min_{\mathbf{P}>0}\left|\mathbf{\Lambda} \mathbf{P} - \mathbf{P}_{\mathrm{noisy}}\right|,
\end{align}
and iterative Bayesian unfolding, i.e., using an iterative formula to update the estimate of $\mathbf{P}_{\mathrm{ideal}}$ within a Bayesian framework~\cite{Nachman2020}.
Notably, measuring local observables from a large quantum system can be reduced to measuring few-qubit subsystems, thus still falling into this category.

\subsubsection{Non-local observables}
As for the large-scale measurements, describing the complete distribution $\mathbf{P}_{\mathrm{noisy}}$ is infeasible due to the exponential scaling.
Instead, one only obtains $M$ samples (generally a constant independent of $N$) from experiments, represented by $M$ bit strings $\{\mathbf{x}^{(m)}_{\rm noisy}\}_{m=1}^M$.

Some applications rely on expectation values of certain local or global observables, for which noisy expectation values can be estimated from these samples
\begin{align}
    \langle O \rangle_{\mathrm{noisy}} = \sum_{\mathbf{x}}\mathbf{P}_{\rm noisy}(\mathbf{x})O(\mathbf{x})\approx \frac{1}{M} \sum_{m=1}^M O (\mathbf{x}^{(m)}_{\rm noisy}),
\end{align}
where experimental samples $\{\mathbf{x}^{(m)}_{\rm noisy}\}_{m=1}^M$ satisfy the probability distribution $\mathbf{P}_{\rm noisy}$.
Here the observable $O$ is diagonal in the computational basis, which is always the case since to measure a non-diagonal observable in real experiments, one needs to first apply a unitary transformation before the measurement.
In this scenario, the ideal expectation values can be efficiently estimated given the readout error matrix $\mathbf{\Lambda}$ as follows.
Firstly, the expectation value can be expressed as
\begin{align}
    \langle O\rangle_{\mathrm{ideal}} = \sum_{\mathbf{y}}\mathbf{P}_{\rm ideal}(\mathbf{y})O(\mathbf{y}),
\end{align}
where $\mathbf{P}_{\rm ideal}(\mathbf{y})$ for a concrete $\mathbf{y}$ can be expanded using Eq.~\eqref{Equ: Inverse} as
\begin{align}
    \mathbf{P}_{\rm ideal}(\mathbf{y}) = \sum_{\mathbf x}\left(\mathbf{\Lambda}^{-1}\right)_{\mathbf{y}, \mathbf{x}}\mathbf{P}_{\rm noisy}(\mathbf{x}).
\end{align}
Similarly, we use experimental samples $\{\mathbf{x}^{(m)}_{\rm noisy}\}_{m=1}^M$ to replace the distribution $\mathbf{P}_{\rm noisy}$ as
\begin{align}
    \mathbf{P}_{\rm ideal}(\mathbf{y}) = \frac{1}{M}\sum_{m=1}^M\left(\mathbf{\Lambda}^{-1}\right)_{\mathbf{y}, \mathbf{x}_{\rm noisy}^{(m)}}.
\end{align}
Finally, the ideal expectation value is expressed as~\cite{Bravyi2021}
\begin{align}
    \langle O \rangle_{\mathrm{ideal}} = \frac{1}{M} \sum_{m=1}^M \left[\sum_{\mathbf{y}} O(\mathbf{y}) \left(\mathbf{\Lambda}^{-1}\right)_{\mathbf{y}, \mathbf{x}^{(m)}_{\rm noisy}}\right].
    \label{Equ: nonlocal}
\end{align}
In general, once this matrix multiplication can be efficiently computed, the readout errors can be mitigated, as demonstrated in both the cases of tensor-product and correlated Markovian readout error models~\cite{Bravyi2021}.

\subsection{Global sampling problem}
Other problems of interest focus on the sampling problem from quantum systems, i.e., generating samples from the ideal distribution $\mathbf{P}_{\mathrm{ideal}}$.
In other words, the goal of REM is to generate $M^{\prime}$ samples $\{\mathbf{y}^{(m)}_{\rm ideal}\}_{m=1}^{M^{\prime}}$ from $\mathbf{P}_{\mathrm{ideal}}$ given the samples $\{\mathbf{x}^{(m)}_{\rm noisy}\}_{m=1}^M$ from $\mathbf{P}_{\mathrm{noisy}}$.
If the probability distribution exhibits certain structures, e.g., some bit strings have much higher probabilities than others such that they always appear in the noisy samples, one can solve the linear equations only in the $M\times M$ subspace spanned by noisy samples as long as concrete matrix elements of $\mathbf{\Lambda}$ can be efficiently calculated~\cite{Nation2021}.
The authors demonstrated this method on GHZ states, where only $00\cdots0$ and $11\cdots1$ bit strings have significant probabilities in the ideal distribution.

This method has two limitations.
First, it relies on the assumption that the noisy samples cover all significant bit strings in the ideal distribution, which may not hold in general.
Particularly, in the famous problem of random circuit sampling~\cite{Arute2019, Wu2021}, the measurement outcomes sampled from a deep random circuit follow the Porter-Thomas distribution, which arises as a direct manifestation of the concentration of measure phenomenon~\cite{Hayden2006, Harrow2009, Aaronson2017}.
In this case, probabilities for most bitstrings cluster around the mean value $1/2^N$, making it impossible to cover all significant bit strings with only constant or polynomially many noisy samples.

Secondly, so far we have focused on the outcomes from a specific measurement basis with many samples.
However, in many applications such as learning quantum systems~\cite{Gebhart2023}, one often needs to measure the quantum state in different bases but with only a few samples (even one sample) for each basis.
A representative example is the classical shadow protocol~\cite{Huang2020}, where the quantum state is measured in randomly chosen bases (e.g., Pauli bases) to construct a classical representation and extract desired properties from the quantum state.
Therefore, it is crucial to extend the REM to these scenarios.

\section{Tensor network characterization of readout errors}
\begin{figure*}
    \centering
    \includegraphics[width=0.9\linewidth]{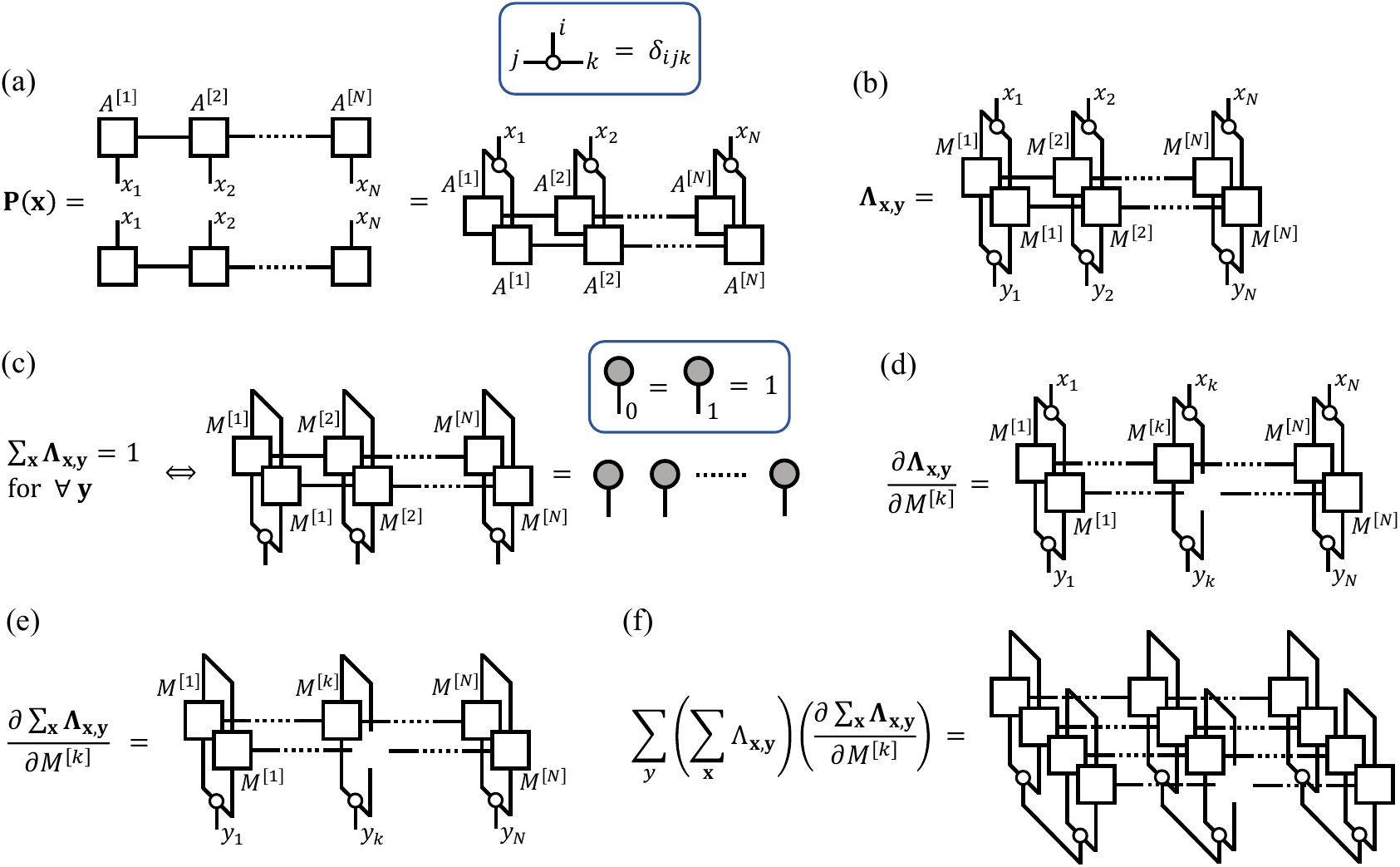}
    \caption{Tensor network representation of classical probability distribution $\mathbf{P}(\mathbf{x})$ and the readout error matrix $\mathbf{\Lambda}_{\mathbf{x},\mathbf{y}}$.
    (a) MPS representation of $\mathbf{P}(\mathbf{x})$ with bond dimension $D$.
    We further introduce a projector to each local site (white circle), realized by a Kronecker delta tensor, to extract the diagonal term and neglect the off-diagonal terms that are irrelevant.
    (b) MPO representation of $\mathbf{\Lambda}_{\mathbf{x},\mathbf{y}}$ with bond dimension $\chi$. 
    Physical indices of each two local tensors $M^{[k]}$ are combined and projected to diagonal terms, where $x_k$ and $y_k$ correspond to the output and input bit values of the $k$-th qubit, respectively.
    (c) $\sum_{\mathbf{x}} \mathbf{\Lambda}_{\mathbf{x},\mathbf{y}}$ can be efficiently calculated by contracting the output physical indices of two local tensors.
    Consequently, the normalization condition $\sum_{\mathbf{x}} \mathbf{\Lambda}_{\mathbf{x},\mathbf{y}} = 1$ means that the resulting tensor network corresponds to a $2^N$-dim unity vector, which can be efficiently represented as an MPS (grey circles) with bond dimension 1.
    (d) Tensor network representation of the gradient $\partial \mathbf{\Lambda}_{\mathbf{x}, \mathbf{y}} / \partial M^{[k]}$.
    (e) Tensor network representation of the gradient $\partial \sum_{\mathbf{x}} \mathbf{\Lambda}_{\mathbf{x},\mathbf{y}} / \partial M^{[k]}$.
    (f) Tensor network representation of the gradient $\sum_{\mathbf{y}}\left(\sum_{\mathbf{x}}\mathbf{\Lambda}_{\mathbf{x}, \mathbf{y}}\right)\left(\partial \sum_{\mathbf{x}} \mathbf{\Lambda}_{\mathbf{x},\mathbf{y}} / \partial M^{[k]}\right)$, which is realized by the inner product of the tensor networks in (c) and (e).}
    \label{Fig: MPO}
\end{figure*}
Generally, the readout error matrix $\mathbf{\Lambda}$ should be short-range correlated, since the crosstalk errors among qubits usually occur locally due to the physical layout of qubits in quantum devices.
Notably, $\mathbf{\Lambda}$ represents a completely classical process acting on measurement outcomes, which can be understood as a conditional probability distribution $\mathbf{P}(\mathbf{x} | \mathbf{y})=\mathbf{\Lambda}_{\mathbf{x},\mathbf{y}}$.
Inspired by previous works that used matrix product states (MPS) to learn classical probability distributions with correlations~\cite{Stoudenmire2016, Han2018}, we propose to use an MPO to represent the readout error matrix $\mathbf{\Lambda}$, which naturally captures the local and short-range correlations among qubits.

\subsection{Tensor network representation of readout error matrix}
Specifically, these MPS-based learning methods use the probability (instead of amplitude) of an MPS with bond dimension $D$ to encode the classical probability distribution, i.e.,
\begin{align}
    \mathbf{P}(\mathbf{x}) = \left|\psi(\mathbf{x})\right|^2 = \left|\Tr\left(A^{[1]}_{x_1} A^{[2]}_{x_2} \cdots A^{[N]}_{x_N}\right)\right|^2,
\end{align}
as shown in Fig.~\ref{Fig: MPO}(a).
This parameterization not only aligns well with the physical interpretation of MPS, i.e., the samples to learn from are generated by measuring a hidden quantum state, but also guarantees the positivity of the probability distribution.
The normalization condition $\sum_{\mathbf{x}} \mathbf{P}(\mathbf{x}) = \braket{\psi|\psi} = 1$ can be efficiently imposed during the optimization process by utilizing the canonical form of MPS~\cite{PerezGarcia2007, Schollwoeck2011}.
Notably, a real-valued MPS is sufficient to represent any classical probability distribution without loss of generality.

Similarly, we can use an MPO with finite bond dimension $\chi$ to represent the readout error matrix $\mathbf{\Lambda}$ as
\begin{align}
    \mathbf{\Lambda}_{\mathbf{x},\mathbf{y}} = |\mathbf{M}_{\mathbf{x}, \mathbf{y}}|^2 =  \left|\Tr\left(M^{[1]}_{x_1,y_1} M^{[2]}_{x_2,y_2} \cdots M^{[N]}_{x_N,y_N}\right)\right|^2,
\end{align}
as shown in Fig.~\ref{Fig: MPO}(b), which also ensures the positivity of each matrix element $\mathbf{\Lambda}_{\mathbf{x},\mathbf{y}}$.
Meanwhile, the normalization condition now becomes $\sum_{\mathbf{x}} \mathbf{\Lambda}_{\mathbf{x},\mathbf{y}} = 1$ for any $\mathbf{y}$, which requires the diagonal elements of $\mathbf{M}^{\dagger}\mathbf{M}$ to be unity, i.e.,
\begin{align}
    \sum_{\mathbf{x}} \mathbf{\Lambda}_{\mathbf{x},\mathbf{y}} = \sum_{\mathbf{x}} \mathbf{M}^{\dagger}_{\mathbf{y}, \mathbf{x}} \mathbf{M}_{\mathbf{x}, \mathbf{y}} = \left(\mathbf{M}^{\dagger}\mathbf{M}\right)_{\mathbf{y}, \mathbf{y}} = 1,\, \forall \mathbf{y},
\end{align}
but with no constraints on the off-diagonal elements, as shown in Fig.~\ref{Fig: MPO}(c).
This is a weaker condition than matrix product unitary (MPU)~\cite{Cirac2017}.

On the other hand, since we are studying a completely classical process, we can also choose a real-valued MPO without loss of generality.
Under this convention, uncorrelated readout errors can be represented by an MPO error model with $\chi=1$, where each local tensor is defined as
\begin{align}
    M^{[k]}_{x_k,y_k} = \sqrt{\mathbf{\Lambda}^{[k]}_{x_k,y_k}}.
\end{align}
For example, consider a single qubit with bit-flip probability $p_{1}$, whose readout error matrix is
\begin{align}
    \mathbf{\Lambda}^{[k]} = \begin{bmatrix}
        1 - p_{1} & p_{1}\\
        p_{1} & 1 - p_{1}
    \end{bmatrix}.
\end{align}
Its corresponding local tensor in the MPO representation is
\begin{align}
    M^{[k]} = \begin{bmatrix}
        \sqrt{1 - p_{1}} & \sqrt{p_{1}} \\
        \sqrt{p_{1}} & \sqrt{1 - p_{1}}
    \end{bmatrix}.
\end{align}
It can be verified that this tensor satisfies the normalization condition by directly calculating
\begin{align}
    \left(M^{[k]}\right)^{\rm T} M^{[k]} = \begin{bmatrix}
        1 & \alpha \\
        \alpha & 1
    \end{bmatrix},
\end{align}
with off-diagonal elements $\alpha = 2\sqrt{p_{1}(1 - p_{1})}$ generally non-zero, though these off-diagonal elements carry little physical significance.

\subsection{Learning the MPO representation of readout errors}
In this section, we discuss how to efficiently learn the MPO representation of the readout error matrix $\mathbf{\Lambda}$ from experimental data.
Inspired by the quantum process tomography method based on tensor network representation~\cite{Torlai2023}, we can prepare and measure a set of $N$-qubit product states $\ket{\mathbf{y}^{(m)}} = \ket{y_1^{(m)}} \otimes \ket{y_2^{(m)}} \otimes \cdots \otimes \ket{y_N^{(m)}}$ randomly chosen from the computational basis, and collect the measurement outcomes $\{\mathbf{x}^{(m)}\}_{m=1}^M$ for each input state $\ket{\mathbf{y}^{(m)}}$.
Then, we can optimize the local tensors of the MPO by minimizing the negative log-likelihood function
\begin{align}
    \mathbb{L} = -\frac{1}{M}\sum_{m=1}^M \log \mathbf{\Lambda}_{\mathbf{x}^{(m)}, \mathbf{y}^{(m)}}
\end{align}
with respect to the constraint $\sum_{\mathbf{x}} \mathbf{\Lambda}_{\mathbf{x},\mathbf{y}} = 1$ for any $\mathbf{y}$.
However, since this normalization condition cannot directly apply to local tensors, we need to impose it by adding a penalty term to the loss function and manually dividing the probability $\mathbf{\Lambda}_{\mathbf{x}^{(m)}, \mathbf{y}^{(m)}}$ by a normalization factor, i.e.,
\begin{align}
    \mathbb{L} = -\frac{1}{M}\sum_{m=1}^M \log \frac{\mathbf{\Lambda}_{\mathbf{x}^{(m)}, \mathbf{y}^{(m)}}}{\sum_{\mathbf{x}} \mathbf{\Lambda}_{\mathbf{x},\mathbf{y}^{(m)}}} + \Delta \sum_{\mathbf{y}} \left(\sum_{\mathbf{x}} \mathbf{\Lambda}_{\mathbf{x},\mathbf{y}} - 1\right)^2,\label{Equ: KL}
\end{align}
where $\Delta$ is a hyperparameter to control the strength of the penalty term.
Throughout this paper, $\Delta$ is set to $1$ for all numerical simulations and experiments, which is sufficient to ensure the normalization condition is satisfied with high precision.

Subsequently, the local tensors can be optimized by gradient-based methods, where the gradient of the loss function $\mathbb{L}$ with respect to each local tensor $M^{[k]}$ reads as
\begin{align}
\begin{aligned}
    \frac{\partial \mathbb{L}}{\partial M^{[k]}} &= \frac{1}{M}\sum_{m=1}^M \left(-\frac{1}{\mathbf{\Lambda}_{\mathbf{x}^{(m)}, \mathbf{y}^{(m)}}} \frac{\partial \mathbf{\Lambda}_{\mathbf{x}^{(m)}, \mathbf{y}^{(m)}}}{\partial M^{[k]}} + \frac{1}{\sum_{\mathbf{x}} \mathbf{\Lambda}_{\mathbf{x},\mathbf{y}^{(m)}}} \frac{\partial \sum_{\mathbf{x}} \mathbf{\Lambda}_{\mathbf{x},\mathbf{y}^{(m)}}}{\partial M^{[k]}}\right) \\ 
    &+ 2\Delta \sum_{\mathbf{y}} \left(\sum_{\mathbf{x}} \mathbf{\Lambda}_{\mathbf{x},\mathbf{y}} - 1\right)  \frac{\partial \sum_{\mathbf{x}}\mathbf{\Lambda}_{\mathbf{x},\mathbf{y}}}{\partial M^{[k]}}.
\end{aligned}
\end{align}
Notably, each term constituting $\mathbb{L}$ and $\partial \mathbb{L} / \partial M^{[k]}$ can be efficiently calculated by contracting the corresponding tensor networks, as illustrated in Fig.~\ref{Fig: MPO}(c)-(f), which can be done in polynomial time with respect to $N$ and $\chi$.
In practice, we can use the Adam optimizer~\cite{Kingma2017} to optimize the local tensors, with the hyperparameters set to $\beta_1 = 0.9$, $\beta_2 = 0.999$, and $\eta=0.002$.

\subsection{Experiments and simulations}
\subsubsection{Experiments on real quantum hardware}
\begin{figure*}
    \centering
    \includegraphics[width=\linewidth]{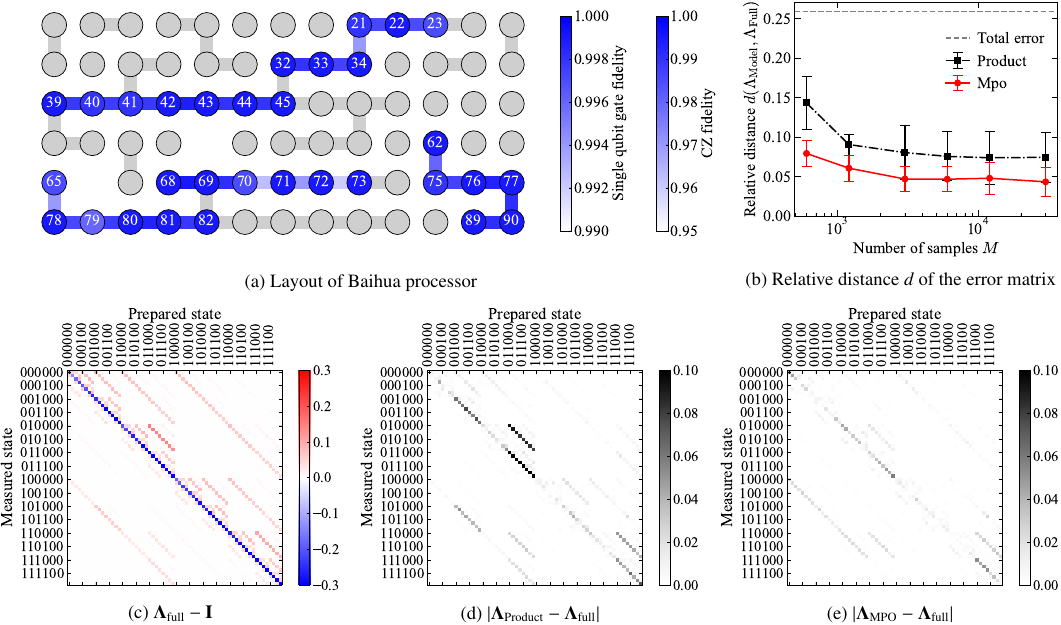}
    \caption{Experiments for readout error characterization on the Baihua superconducting quantum chip.
    (a) The layout of the Baihua superconducting quantum chip.
    Five non-overlapping six-qubit chains with nearest-neighbor connections are selected for experiments.
    (b) Relative distance of the uncorrelated $\mathbf{\Lambda}_{\rm Product}$ and MPO readout error models $\mathbf{\Lambda}_{\rm MPO}$ from the fully characterized readout error matrix $\mathbf{\Lambda}_{\rm Full}$ under different measurement times $M$.
    (c-e) Results on the \{65, 78, 79, 80, 81, 82\} qubit chain under $M=30000$ shots.
    (c) Matrix elements of the fully characterized readout error matrix $\mathbf{\Lambda}_{\rm Full}$ (visualized as $\mathbf{\Lambda}_{\rm Full} - \mathbf{I}$).
    (d) Absolute difference between $\mathbf{\Lambda}_{\rm Product}$ and $\mathbf{\Lambda}_{\rm Full}$.
    (e) Absolute difference between $\mathbf{\Lambda}_{\rm MPO}$ and $\mathbf{\Lambda}_{\rm Full}$.}
    \label{Fig: Exp1}
\end{figure*}

We illustrate the necessity of considering correlated readout errors by performing five parallel experiments on the Baihua superconducting quantum chip from Quafu quantum cloud platform~\cite{Chen2022}.
In these experiments, we select five non-overlapping six-qubit $(N=6)$ chains with nearest-neighbor connections arranged in a two-dimensional (2D) layout, as shown in Fig.~\ref{Fig: Exp1}(a).

In each experiment, we first perform a full characterization of the readout error matrix $\mathbf{\Lambda}_{\rm Full}$ by preparing and measuring all $2^N = 64$ computational basis states, each measured with $5000$ shots, as a reference.
Next, we compare the uncorrelated readout error model and the MPO model.
For the uncorrelated model $\mathbf{\Lambda}_{\rm Product}$, we characterize the readout error matrix $\mathbf{\Lambda}^{[k]}$ for each qubit by preparing and measuring $\ket{0}$ and $\ket{1}$ states with $s$ shots, leading to a total of $M = 2 \times N \times s = 12s$ shots.
To learn the MPO readout error model $\mathbf{\Lambda}_{\rm MPO}$, we prepare and measure $M=12s$ random product states $\ket{\mathbf{y}^{(m)}}$ with single shot for each state, such that the total measurement times are the same as the uncorrelated model.
In our implementation, the MPO bond dimension is set to $\chi = 4$.
The measurement dataset is randomly divided into a training set ($80\%$) and a validation set ($20\%$).
Training is performed for $100$ epochs with a batch size of $256$, with each epoch corresponding to one complete traversal of the entire training dataset.

We investigate the performance of these two models under different measurement times $M$ by evaluating their relative distance from the full error matrix $\mathbf{\Lambda}_{\rm Full}$.
The relative distance of $\mathbf{A}$ from the ideal $\mathbf{A}^{0}$ is defined as
\begin{align}
d(\mathbf{A}, \mathbf{A}^{0}) \equiv \|\mathbf{A} - \mathbf{A}^{0}\|_F / \|\mathbf{A}^{0}\|_F,
\end{align}
where $\|\cdot\|_F$ denotes the Frobenius norm.
As shown in Fig.~\ref{Fig: Exp1}(b), the MPO model converges to $d(\mathbf{\Lambda}_{\rm MPO}, \mathbf{\Lambda}_{\rm Full}) \approx 0.043$, achieving a smaller relative distance than the uncorrelated model that converges to $d(\mathbf{\Lambda}_{\rm Product}, \mathbf{\Lambda}_{\rm Full}) \approx 0.074$.
Moreover, the MPO model consistently outperforms the uncorrelated model across different measurement times $M$, verifying the necessity to consider correlated readout errors in real experiments.
Nevertheless, compared with the total error strength characterized by $d(\mathbf{I}, \mathbf{\Lambda}_{\rm Full})\approx 0.26$, which is plotted as a dashed line, both models have successfully captured most of the readout errors.
Figs.~\ref{Fig: Exp1}(c)-(e) present the detailed matrix elements of $\mathbf{\Lambda}_{\rm Full}$ and the absolute differences between $\mathbf{\Lambda}_{\rm Product}$, $\mathbf{\Lambda}_{\rm MPO}$ and $\mathbf{\Lambda}_{\rm Full}$ for the experiment performed on the \{65, 78, 79, 80, 81, 82\} qubit chain.
Both models are learned under $M=30000$ shots.
Notably, the MPO model significantly reduces the absolute error across almost all matrix elements compared to the uncorrelated model, demonstrating its superior capability in capturing correlated readout errors.

\subsubsection{Numerical simulations for larger systems}
\begin{figure*}
    \centering
    \includegraphics[width=\linewidth]{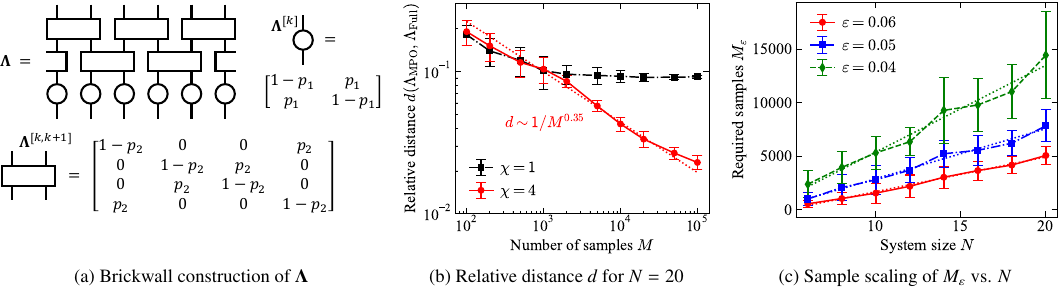}
    \caption{Numerical simulations for larger systems up to $N=20$ qubits.
    (a) Brickwall construction for the readout error matrix $\mathbf{\Lambda}$, where single-qubit error matrices $\mathbf{\Lambda}^{[k]}$ are placed on each qubit, and two-qubit error matrices $\mathbf{\Lambda}^{[k, k+1]}$ are placed on each nearest-neighbor pair of qubits.
    (b) Relative distance of the MPO readout error models $\mathbf{\Lambda}_{\rm MPO}$ from the full error matrix $\mathbf{\Lambda}_{\rm Full}$ for $N=20$.
    (c) Number of samples $M$ required for the MPO model to achieve a relative distance $d(\mathbf{\Lambda}_{\rm MPO}, \mathbf{\Lambda}_{\rm Full}) < \varepsilon $ under different system sizes $N$.}
    \label{Fig: Sim1}
\end{figure*}
To further demonstrate the scalability of our method, we perform numerical simulations for larger systems up to $N=20$ qubits.
In these simulations, we assume a brickwall structure for the readout error matrix $\mathbf{\Lambda}_{\rm Full}$, where single-qubit error matrices
\begin{align}
    \mathbf{\Lambda}^{[k]} = \begin{bmatrix}
        1 - p_{1} & p_{1}\\
        p_{1} & 1 - p_{1}
    \end{bmatrix},\, p_{1} = 0.03,
\end{align}
are placed on each qubit, followed by two-qubit error matrices
\begin{align}
    \mathbf{\Lambda}^{[k, k+1]} = \begin{bmatrix}
        1 - p_{2} & 0 & 0 & p_{2}\\
        0 & 1 - p_{2} & p_{2} & 0\\
        0 & p_{2} & 1 - p_{2} & 0\\
        p_{2} & 0 & 0 & 1 - p_{2}
    \end{bmatrix},\, p_{2} = 0.005,
\end{align}
that are placed on each nearest-neighbor pair of qubits to model simultaneous flipping of both bits due to crosstalk effects, as shown in Fig.~\ref{Fig: Sim1}(a).
For $N=6$, the relative distance of this full error matrix from the identity matrix is $d(\mathbf{I}, \mathbf{\Lambda}_{\rm Full}) \approx 0.24$, which is comparable to previous experimental results.

Next, we learn the MPO readout error models $\mathbf{\Lambda}_{\rm MPO}$ with different bond dimensions from $M$ random samples, and evaluate their relative distances from the full error matrix $\mathbf{\Lambda}_{\rm Full}$.
As shown in Fig.~\ref{Fig: Sim1}(b) for $N=20$, larger sample numbers $M$ generally lead to smaller relative distances for both $\chi=1$ and $\chi=4$ models.
However, the $\chi=1$ model, which corresponds to the uncorrelated readout error model, fails to capture the correlated errors in $\mathbf{\Lambda}_{\rm Full}$, resulting in an early saturation of $d$ as $M$ increases.
On the contrary, the $\chi=4$ model continues to improve as $M$ increases, consistently outperforms the $\chi=1$ model for $M> 1000$, and eventually achieves a relative distance $d\approx 0.02$ with $M=10^5$ samples, which is significantly smaller than the total number of parameters in $\mathbf{\Lambda}_{\rm Full}$, i.e., $2^{2N} = 2^{40}$.
The scaling behavior between $d$ and $M$ is fitted as $d \sim M^{-0.35}$, implying a power-law improvement with the increase of sample number.

Finally, we investigate the scalability of our method by examining the required sample number $M$ to achieve a certain relative distance $\varepsilon$, i.e., $d(\mathbf{\Lambda}_{\rm MPO}, \mathbf{\Lambda}_{\rm Full}) < \varepsilon$, under different system sizes $N$.
As shown in Fig.~\ref{Fig: Sim1}(c), the required sample number $M$ shows a nearly linear scaling with $N$ for different thresholds $\varepsilon$, which is significantly more efficient than the exponential scaling of the full characterization method.
All numerical simulations are performed $50$ times, and the mean values are plotted with error bars representing the standard deviations.
In summary, our method not only captures the correlated readout errors in real experiments, but also scales efficiently to larger systems with a sample complexity that grows only near-linearly with the system size.

In the following sections, we focus on how to mitigate readout errors using the learned MPO model $\mathbf{\Lambda}_{\rm MPO}$ in various applications.
Specifically, we will demonstrate that the MPO model can be directly integrated into existing REM methods that are originally designed for uncorrelated models.
Moreover, we will discuss several new applications of REM beyond known methods that can be efficiently solved with the MPO model, especially the global sampling problems for random circuits or under random measurements, as well as integrations with QEC decoding.

\section{REM of nonlocal observables}
Since the readout errors in local probability distributions or observables can be efficiently mitigated by a full characterization of the marginal readout error matrix, we directly focus on the mitigation of readout errors in nonlocal observables.
We will show that, with the learned MPO model $\mathbf{\Lambda}_{\rm MPO}$, the ideal expectation value of a nonlocal observable can be efficiently estimated from the noisy samples by directly applying the formula in Eq.~\eqref{Equ: nonlocal}, where the matrix multiplication can be efficiently computed by contracting the corresponding tensor networks.
Therefore, the main challenges lie in the MPS representation of $O(\mathbf{y})$, which is viewed as a vector in the computational basis, and the efficient calculation of the error matrix inverse $\mathbf{\Lambda}_{\rm MPO}^{-1}$.

\subsection{Tensor-network representation of quasi-probability distribution}
\begin{figure}
    \centering
    \includegraphics[width=0.87\linewidth]{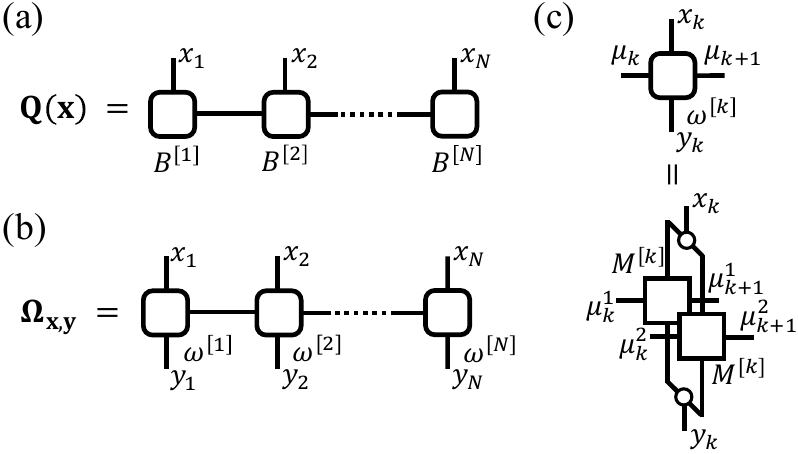}
    \caption{Tensor network representation of the (conditional) quasi-probability distribution.
    (a) Single-layer MPS representation of $\mathbf{Q}(\mathbf{x})$.
    (b) Single-layer MPO representation of $\mathbf{\Omega}_{\mathbf{x},\mathbf{y}}$.
    (c) Transformation from the double-layer representation to the single-layer one by grouping the virtual indices and projecting the physical indices to diagonal terms.}
    \label{Fig: Single}
\end{figure}
Before discussing the applications of REM, we first introduce another type of tensor network representation of classical (conditional) probability distributions.
Instead of the double-layer encoding as used in Fig.~\ref{Fig: MPO}(a) and (b), we can directly use single-layer MPS and MPO to represent the probability distribution $\mathbf{P}(\mathbf{x})$ and the conditional probability distribution $\mathbf{Q}(\mathbf{x} | \mathbf{y}) = \mathbf{\Omega}_{\mathbf{x},\mathbf{y}}$, as shown in Fig.~\ref{Fig: Single}(a) and (b), respectively.
This single-layer representation is more straightforward and easier to implement, but it does not guarantee the positivity of the probability distribution.
Therefore, it is more suitable for representing the quasi-probability distribution that may contain negative entries, which often appears in the inverse of the readout error matrix $\mathbf{\Lambda}^{-1}$, as well as the error-mitigated distributions $\mathbf{\Lambda}^{-1}\mathbf{P}_{\mathrm{noisy}}$ when the number of samples is small and thus subject to large statistical fluctuations~\cite{Nachman2020}.

To transform the double-layer representation to the single-layer one, one can simply group the virtual indices of the local tensors, and project the physical indices to diagonal terms, i.e.,
\begin{align}
    \left(\omega^{[k]}\right)_{x_k, y_k}^{\mu_{k}, \mu_{k+1}} = \left(M^{[k]}\right)_{x_k, y_k}^{\mu_{k}^1, \mu_{k+1}^1} \left(M^{[k]}\right)_{x_k, y_k}^{\mu_{k}^2, \mu_{k+1}^2},
\end{align}
with $\mu_{k} = \left\{\mu_{k}^1, \mu_{k}^2\right\}$ representing the grouped virtual indices, as shown in Fig.~\ref{Fig: Single}(c).
After this transformation, the resulting single-layer representation can be directly used to perform REM in various applications, as discussed in the following sections.

\subsection{Error mitigation with tensor contraction}
\begin{figure}
    \centering
    \includegraphics[width=\linewidth]{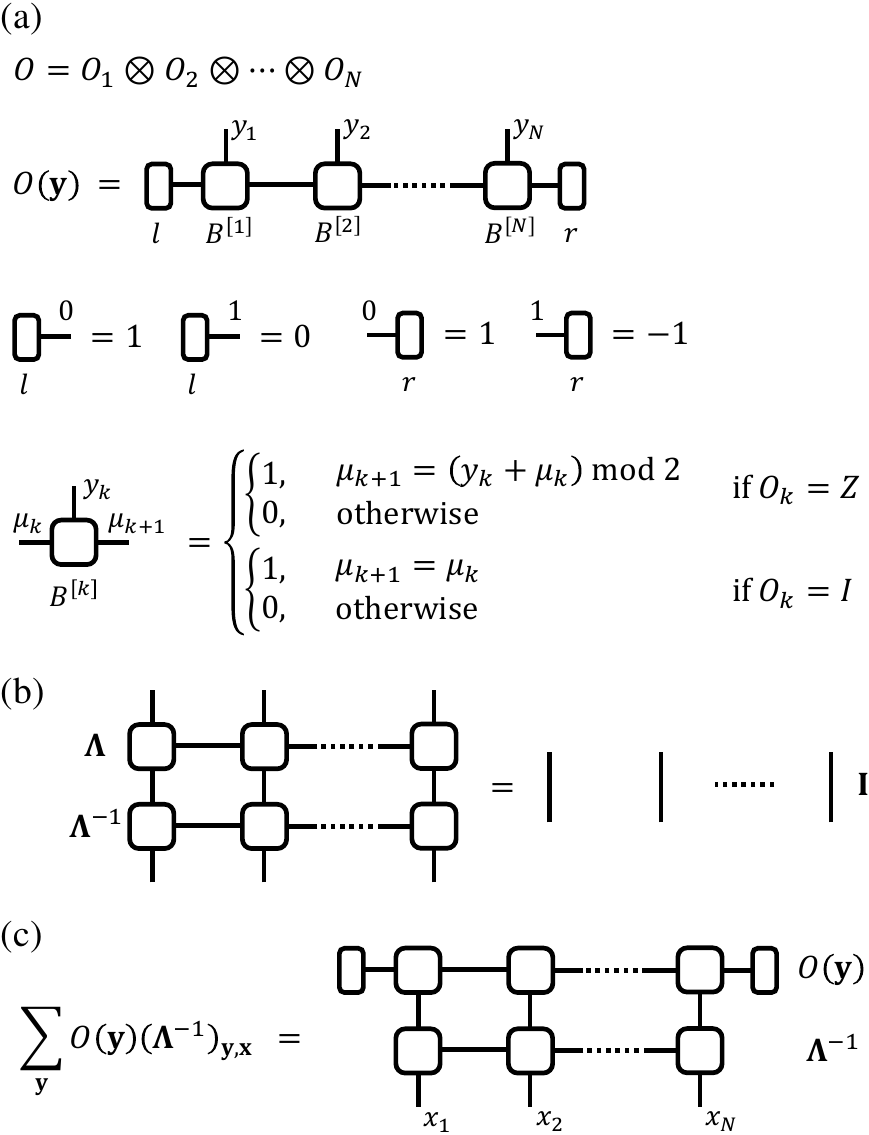}
    \caption{Tensor network representation of the REM for nonlocal observables.
    (a) MPS representation of $O(\mathbf{y})$ with bond dimension $D=2$.
    (b) MPO representation of $\mathbf{\Lambda}^{-1}_{\mathbf{y}, \mathbf{x}}$.
    (c) Tensor network representation of the matrix multiplication $\sum_{\mathbf{y}} O(\mathbf{y}) \left(\mathbf{\Lambda}^{-1}\right)_{\mathbf{y}, \mathbf{x}}$, which can be efficiently calculated by contracting the corresponding tensor networks.}
    \label{Fig: Nonlocal}
\end{figure}
Firstly, without loss of generality, we can assume that the observable $O$ is a product of Pauli operators, i.e., $O = O_1 \otimes O_2 \otimes \cdots \otimes O_N$, where each $O_k$ is a single-qubit Pauli operator (including the identity $I$).
As discussed before, to measure a non-diagonal observable in real experiments, one needs to first apply a unitary transformation before the measurement.
Therefore, we can focus on the case where $O$ is only composed of $Z$ and $I$ operators.
In this case, $O(\mathbf{y})$ counts the parity of the bits in $\mathbf{y}$ that correspond to $Z$ operators, which can be efficiently represented as an MPS with bond dimension $D=2$ in Fig.~\ref{Fig: Nonlocal}(a).
Starting from the left boundary condition of $\mu_1=0$, the virtual bond carries the parity information of the bits from the leftmost site to the current site, which is flipped if the current site corresponds to a $Z$ operator and the bit value is $1$, and remains unchanged otherwise.
The right boundary condition is then used to project the final parity information to the expectation value of the observable.

Secondly, the error matrix inverse $\mathbf{\Lambda}^{-1}$ can be efficiently calculated by using the variational method proposed in Ref.~\cite{Guo2022}.
Specifically, we find another single-layered MPO $\mathbf{\Omega}$ that minimizes
\begin{align}
\begin{aligned}
    \|\mathbf{\Lambda} \mathbf{\Omega} - \mathbf{I}\|_F^2 &= \Tr\left(\mathbf{\Omega}^{\rm T} \mathbf{\Lambda}^{\rm T} \mathbf{\Lambda} \mathbf{\Omega}\right)\\
     &- 2\Tr\left(\mathbf{\Omega}\mathbf{\Lambda}\right) + \Tr\left(\mathbf{I}\right),
\end{aligned}
\end{align}
where $\mathbf{I}$ is the identity matrix, as shown in Fig.~\ref{Fig: Nonlocal}(b).
This loss function is quadratic with respect to each local tensor in $\mathbf{\Omega}$, and thus can be efficiently optimized by a sweeping algorithm similar to the density matrix renormalization group (DMRG) method~\cite{Schollwoeck2011}.
After obtaining the optimal $\mathbf{\Omega}$, we can directly use it as the error matrix inverse $\mathbf{\Lambda}^{-1}$ to perform REM for nonlocal observables by contracting the corresponding tensor networks, as shown in Fig.~\ref{Fig: Nonlocal}(c).

\subsection{Experiments and simulations}
\begin{figure*}
    \includegraphics[width=\linewidth]{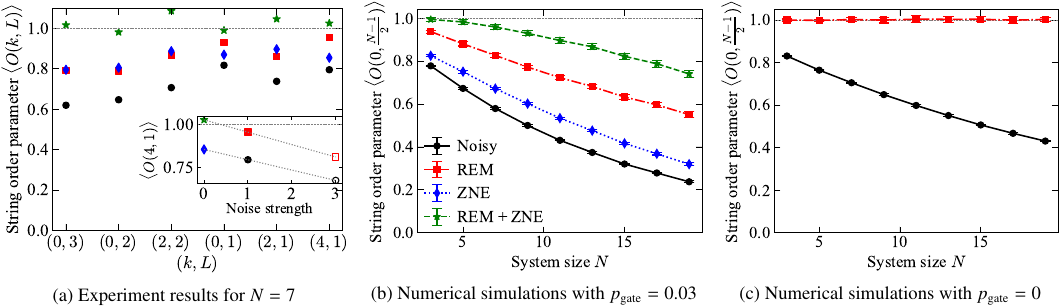}
    \caption{REM for nonlocal string order parameters for 1D cluster states. Three subfigures share the same legend.
    (a) Experiments on the Baihua superconducting quantum chip for $N=7$.
    The insets illustrate the procedure of combining ZNE and REM to extract the ideal expectation values of nonlocal observables.
    (b) Numerical simulations for larger systems up to $N=19$ qubits with gate noise $p_{\rm gate}=0.03$.
    (c) Numerical simulations for larger systems up to $N=19$ qubits with $p_{\rm gate}=0$.}
    \label{Fig: Cluster}
\end{figure*}
As a concrete example, we consider the 1D cluster state
\begin{align}
    \ket{\psi_{\rm Cluster}} = \prod_{k=1}^{N-1} {\rm CZ}_{k, k+1} \ket{+}^{\otimes N},
\end{align}
where ${\rm CZ}_{k, k+1}$ is the controlled-Z gate acting on the $k$-th and $(k+1)$-th qubits.
This state has many interesting properties, e.g., it is a resource state for measurement-based quantum computation~\cite{Raussendorf2001}, and it also belongs to a symmetry-protected topological phase protected by the $\mathbb{Z}_2 \times \mathbb{Z}_2$ symmetry generated by $\prod_{k\in \mathrm{odd}} X_k$ and $\prod_{k\in \mathrm{even}} X_k$~\cite{Chen2011}.
It is stabilized by the operators
\begin{align}
    h_k = Z_{k-1} X_k Z_{k+1} \quad \text{for } k=2, 3, \cdots, N-1,
\end{align}
leading to a long-range string order characterized by the nonlocal observable
\begin{align}
    O(k, L) \equiv Z_{k} X_{k+1} X_{k+3} \cdots X_{k+2L-1} Z_{k+2L} = \prod_{l=1}^{L} h_{k+2l-1},
\end{align}
which is the product of $L$ consecutive stabilizers.

We first prepare a $N=7$ cluster state on the Baihua superconducting quantum chip using \{39, 40, 41, 42, 43, 44, 45\} qubits.
Next, we measure non-local observables $O(k, L)$ up to $L=3$ with even $k$, which can be simultaneously measured in one round of experiments under the staggered basis of $Z_0X_1Z_2X_3Z_4X_5Z_6$ with $50000$ shots.
Note that $O(k,L)$ acts as Pauli $Z$ on its two (even) endpoints and Pauli $X$ on the odd interior sites, while the even interior sites $k+2, k+4, \cdots, k+2L-2$ carry the identity operator.
Since the identity is compatible with any measurement basis, these sites can be measured in an arbitrary basis and their outcomes are simply discarded.
Therefore, measuring every qubit in the staggered basis, i.e., the even (odd) sites in the $Z$ ($X$) basis, is sufficient to evaluate all $O(k,L)$ with even $k$ simultaneously.
To mitigate the two-qubit gate errors during the state preparation, we apply the zero-noise extrapolation (ZNE) method~\cite{Temme2017, Li2017} by inserting additional pairs of CZ gates (thereby effectively amplifying the noise strength by a factor of 3) and extrapolating the expectation values to the zero-noise limit.
We then learn the MPO readout error model $\mathbf{\Lambda}_{\rm MPO}$ with bond dimension $\chi=4$ from $M=50000$ random samples, as illustrated before.
Finally, we evaluate the expectation values of these observables $O(k, L)$ with and without REM by using the learned MPO model $\mathbf{\Lambda}_{\rm MPO}$.
The experimental results are shown in Fig.~\ref{Fig: Cluster}(a), where the procedure of combining ZNE and REM to extract the ideal expectation values is illustrated in the insets using the example of $O(4, 1)$.
We can see that the REM significantly improves the estimation of the ideal expectation values, especially for larger $L$ where the observable is more nonlocal and thus more susceptible to readout errors.
Nevertheless, the mitigated expectation values seem to have a larger uncertainty around the ideal value, which is attributed to negative entries in the error matrix inverse $\mathbf{\Lambda}_{\rm MPO}^{-1}$, where the statistical fluctuations are amplified by these negative entries, leading to a larger variance of the mitigated expectation values~\cite{Nachman2020,Takagi2022}.

Meanwhile, we perform numerical simulations for larger systems up to $N=19$ qubits, targeting the longest string order parameter $O(0, \frac{N-1}{2})$ for each system size $N$.
We consider the same readout error model as used in Fig.~\ref{Fig: Sim1}(a), accompanied by a depolarizing noise model with gate error probability $p_{\rm gate}=0.03$ for CZ gates during the cluster state preparation.
The number of shots for measuring the noisy expectation values and learning the MPO readout error model are both set to $M=50000$.
The results are shown in Fig.~\ref{Fig: Cluster}(b), where for $N=7$ the noisy result $\braket{O(0, 3)}\approx 0.58$ is close to the experimental result (0.62) in Fig.~\ref{Fig: Cluster}(a), demonstrating the validity of our noise model.
Generally, the combination of two error mitigation techniques significantly improves the estimation of the ideal expectation values for all system sizes, though with a larger uncertainty similar to the experimental results.
Nevertheless, the mitigated expectation values still deviate from the ideal values, especially for larger system sizes, which might be attributed to the inaccurate estimation of the error matrix inverse $\mathbf{\Lambda}_{\rm MPO}^{-1}$ due to the limited number of samples, as well as the biased nature of the ZNE method~\cite{Cai2023}.

To further divide the contribution of readout errors and gate errors, we perform another set of numerical simulations without gate errors, i.e., $p_{\rm gate}=0$.
The results are shown in Fig.~\ref{Fig: Cluster}(c), where the mitigated expectation values are much closer to the ideal values with a smaller uncertainty, especially for larger system sizes.
This means that a moderate number of samples, e.g., $M=50000$, is sufficient to learn an accurate MPO readout error model and perform effective REM for nonlocal observables, demonstrating the effectiveness of the MPO REM method.

\section{REM of global sampling problems}
The global sampling problems refer to the task of sampling from the ideal output distribution of a quantum state, i.e., $\mathbf{P}_{\rm ideal}(\mathbf{x}) = |\braket{\mathbf{x}|\varphi}|^2$.
This task is of great importance in various applications, such as quantum supremacy experiments~\cite{Arute2019, Wu2021} and quantum machine learning~\cite{Biamonte2017, Benedetti2019, Cerezo2022}.
Notably, not all these tasks require REM.
For instance, in a typical quantum machine learning experiment, a parameterized quantum circuit is trained on samples from a target distribution and then used to generate new samples from the same distribution.
In this case, the readout errors in both the training process and the generation process can be viewed as a gauge transformation of the ideal distribution $\mathbf{P}_{\rm ideal}$, which does not affect the generated new samples.
Therefore, we focus on the sampling problems from a fixed quantum state, where the readout errors can significantly distort the output distribution and thus require REM to recover the ideal distribution $\mathbf{P}_{\rm ideal}$.
The REM for the sampling problems can thus be formulated as follows: given a set of noisy samples $\{\mathbf{x}^{(m)}_{\rm noisy}\}_{m=1}^M$ drawn from the noisy distribution $\mathbf{P}_{\rm noisy} = \mathbf{\Lambda}\mathbf{P}_{\rm ideal}$, how can we efficiently generate new samples $\{\mathbf{y}_{\rm ideal}^{(m)}\}_{m=1}^{M^\prime}$ that are close to the ideal distribution $\mathbf{P}_{\rm ideal}$?

\subsection{MPS modeling of the ideal distribution}
The basic idea is to learn a classical distribution $\mathbf{P}$ by combining the readout error matrix $\mathbf{\Lambda}$, such that $\mathbf{\Lambda}\mathbf{P}$ has the maximum likelihood to generate the observed noisy samples $\{\mathbf{x}^{(m)}_{\rm noisy}\}_{m=1}^M$.
The loss function for this optimization can be defined as
\begin{align}
    \mathbb{L} = -\frac{1}{M}\sum_{m=1}^M \log \left[\left(\mathbf{\Lambda}\mathbf{P}\right)(\mathbf{x}^{(m)}_{\rm noisy})\right],
\end{align}
where the readout error matrix $\mathbf{\Lambda}$ can be first learned from calibration experiments as discussed before.
After learning the optimal distribution $\mathbf{P}$, we can generate new samples $\{\mathbf{y}^{(m)}\}_{m=1}^{M^\prime}$ from $\mathbf{P}$, which are expected to be close to the ideal distribution $\mathbf{P}_{\rm ideal}$.
This formalism is quite general and can be applied to any classical model for the ideal distribution $\mathbf{P}$, such as the restricted Boltzmann machine~\cite{Torlai2020}, the autoregressive model~\cite{Sharir2020}, or even the transformer model~\cite{RocaJerat2024}.

Here, we focus on the double-layered MPS model shown in Fig.~\ref{Fig: MPO}(a) to approximate $\mathbf{P}_{\rm ideal}$~\cite{Han2018} due to its simplicity and natural compatibility with the MPO model for readout errors.
In general, the underlying quantum state $\ket{\varphi}$ can be an arbitrary state, making the classical modeling of $\mathbf{P}_{\rm ideal}$ challenging.
Nevertheless, we could still use an MPS to approximate the ideal distribution $\mathbf{P}_{\rm ideal}$ as close as possible.
If the quantum state $\ket{\varphi}$ itself can be represented by an MPS with low bond dimension, then $\mathbf{P}_{\rm ideal}$ can be modeled by the same MPS.
As for those $\ket{\varphi}$ that cannot be efficiently represented by an MPS, we can still use an MPS with a moderate bond dimension to capture the main features of $\mathbf{P}_{\rm ideal}$ since we are now working with a classical distribution rather than a quantum state, which is generally easier to approximate.
Even if the MPS model cannot faithfully capture the properties of $\mathbf{P}_{\rm ideal}$, it might still be useful if the learned MPS model can provide a better approximation than the noisy distribution $\mathbf{P}_{\rm noisy}$.

Similar to the learning process of the MPO readout error model, we can optimize the local tensors of the MPS by minimizing the negative log-likelihood function $\mathbb{L}$ with respect to the constraint $\sum_{\mathbf{x}} \mathbf{P}(\mathbf{x}) = 1$ for normalization.
The loss function can be modified as
\begin{align}
    \mathbb{L} = -\frac{1}{M}\sum_{m=1}^M \log \frac{\left(\mathbf{\Lambda}\mathbf{P}\right)(\mathbf{x}^{(m)}_{\rm noisy})}{\sum_{\mathbf{y}} \mathbf{P}(\mathbf{y})}.
    \label{Equ: Loss}
\end{align}
The gradient of the loss function with respect to each local tensor $A^{[k]}$ reads as
\begin{align}
\begin{aligned}
    \frac{\partial \mathbb{L}}{\partial A^{[k]}} &= -\frac{1}{M}\sum_{m=1}^M \left[\frac{1}{\left(\mathbf{\Lambda}\mathbf{P}\right)(\mathbf{x}^{(m)}_{\rm noisy})} \frac{\partial \left(\mathbf{\Lambda}\mathbf{P}\right)(\mathbf{x}^{(m)}_{\rm noisy})}{\partial A^{[k]}}\right]\\
    & + \frac{1}{\sum_{\mathbf{y}} \mathbf{P}(\mathbf{y})} \frac{\partial \sum_{\mathbf{y}} \mathbf{P}(\mathbf{y})}{\partial A^{[k]}}.
\end{aligned}
\end{align}
Therefore, this optimization can be efficiently performed by contracting the corresponding tensor networks to calculate the loss function and its gradients.
We use the Adam optimizer with the same hyperparameters as before to optimize the local tensors of the MPS.
Finally, after learning the optimal MPS model $\mathbf{P}_{\rm MPS}$, we can generate new samples $\{\mathbf{y}^{(m)}_{\rm MPS}\}_{m=1}^{M^\prime}$ from $\mathbf{P}_{\rm MPS}$ by sequentially sampling each bit according to the conditional probability distribution~\cite{Han2018}.

\subsection{Experiments and simulations}
\begin{figure*}
    \centering
    \includegraphics[width=0.66\linewidth]{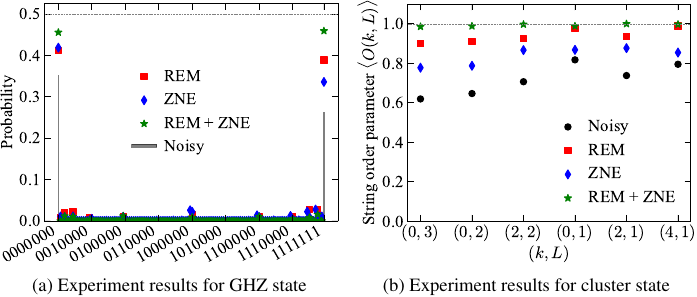}
    \caption{REM for global sampling problems.
    (a) Experiments on the Baihua superconducting quantum chip for a $N=7$ GHZ state.
    (b) Experiments on the Baihua superconducting quantum chip for a $N=7$ cluster state.}
    \label{Fig: Sampling}
\end{figure*}
We first consider the simple case where the ideal quantum state $\ket{\varphi}$ itself is an MPS with low bond dimension, such that the ideal distribution $\mathbf{P}_{\rm ideal}$ can be exactly represented by an MPS.
We adopt the GHZ state $\ket{\varphi} = \frac{1}{\sqrt{2}} (\ket{0}^{\otimes N} + \ket{1}^{\otimes N})$ as an example, which can be represented by an MPS with $D=2$.
We prepare a $N=7$ GHZ state on the Baihua superconducting quantum chip using the same chain as before, and gather the noisy samples $\{\mathbf{x}^{(m)}_{\rm noisy}\}_{m=1}^M$ with $M=50000$ shots.
Using the MPO readout error model $\mathbf{\Lambda}_{\rm MPO}$ with $\chi=4$ learned before, we optimize the MPS model $\mathbf{P}_{\rm MPS}$ with $D=2$ by minimizing the negative log-likelihood function $\mathbb{L}$ in Eq.~\eqref{Equ: Loss}.
Finally, we generate $M^{\prime}=M=50000$ new samples $\{\mathbf{y}^{(m)}_{\rm MPS}\}_{m=1}^{M^{\prime}}$ from the learned MPS model $\mathbf{P}_{\rm MPS}$.

As we are working with a small system size, we can directly evaluate the probability distributions from the noisy samples $\{\mathbf{x}^{(m)}_{\rm noisy}\}_{m=1}^M$ as well as the error-mitigated samples $\{\mathbf{y}^{(m)}_{\rm MPS}\}_{m=1}^{M^{\prime}}$ by counting the frequency of each outcome.
We also combine the ZNE method to mitigate the gate errors during the state preparation by extrapolating each entry of the probability distribution to the zero-noise limit.
Our method of REM guarantees the positivity of the mitigated distribution, while the ZNE method may lead to negative entries due to the extrapolation, which are truncated to zero and renormalized for comparison.
Notably, explicit probability distributions and subsequent ZNE are feasible only for small systems, but are impossible for larger systems due to the exponential growth of the outcome space.
The results in Fig.~\ref{Fig: Sampling}(a) show that the combination of ZNE and REM significantly improves the estimation of the ideal distribution, where the probabilities of $0000000$ and $1111111$ are much closer to the ideal value of $0.5$, while most of the unintended outcomes are significantly suppressed, demonstrating the effectiveness of our method.

Next, we perform the same procedure for a $N=7$ cluster state, where the ideal distribution $\mathbf{P}_{\rm ideal}$ can also be exactly represented by an MPS with $D=2$.
In this case, we estimate the same set of string order parameters as in Fig.~\ref{Fig: Cluster}(a) from the noisy samples $\{\mathbf{x}^{(m)}_{\rm noisy}\}_{m=1}^M$ as well as the error-mitigated samples $\{\mathbf{y}^{(m)}_{\rm MPS}\}_{m=1}^{M^{\prime}}$, and then extrapolate these expectation values to the zero-noise limit by the ZNE method.
The results in Fig.~\ref{Fig: Sampling}(b) show that the mitigated samples provide a much better estimation of the ideal expectation values than the noisy samples.
Moreover, compared to the direct REM for nonlocal observables shown in Fig.~\ref{Fig: Cluster}(a), the sampling-based method can further improve the estimation of the ideal expectation values with a smaller fluctuation.
This improvement is attributed to the fact that the sampling-based method is restricted to the manifold of finite-bond-dimension MPS, which can be viewed as a regularization that suppresses the statistical fluctuations in the error-mitigated distribution.
However, the direct REM for nonlocal observables introduced in the previous section does not rely on the MPS modeling of the ideal distribution, and thus has a broader applicability.

\begin{figure*}
    \centering
    \includegraphics[width=\linewidth]{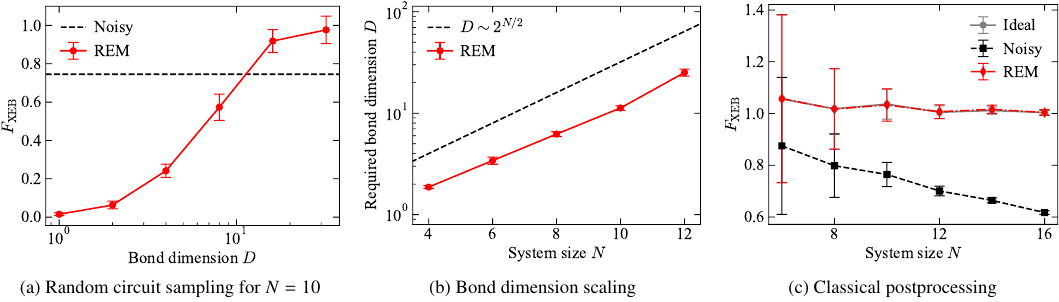}
    \caption{REM for random circuit sampling.
    (a) XEB fidelity $F_{\rm XEB}$ for noisy and REM samples with $N=10$.
    (b) Bond dimension scaling for the crossing between noisy and REM results.
    (c) Classical postprocessing to estimate $F_{\rm XEB}$ without readout errors.}
    \label{Fig: RCS}
\end{figure*}
Finally, we perform numerical simulations for random circuit sampling, where the ideal quantum state $\ket{\varphi}$ is generated by applying a random quantum circuit composed of Haar-random $SU(4)$ gates with depth $d=N$ on the initial state $\ket{0}^{\otimes N}$.
The gates in the circuit are taken to be error-free, so that the readout error is the only noise source in this simulation.
To estimate the quality of the generated samples, we use the cross-entropy benchmarking (XEB) fidelity defined as
\begin{align}
    F_{\rm XEB} = \frac{2^N }{M}\sum_{m} \mathbf{P}_{\rm ideal}(\mathbf{x}^{(m)}) - 1,
\end{align}
where $\mathbf{P}_{\rm ideal}(\mathbf{x})$ is the ideal distribution that needs to be calculated by contracting the corresponding random circuit.
This value is close to zero for random samples, and close to one for ideal samples when the system size is large enough to ensure the anti-concentration of the ideal distribution~\cite{Harrow2017}.
As a result, even the readout-error-free samples do not yield an XEB fidelity of exactly one, and the goal of REM here is therefore to recover the readout-error-free XEB value from the noisy samples.

We consider the same readout error model as before, and gather $M=50000$ noisy samples $\{\mathbf{x}^{(m)}_{\rm noisy}\}_{m=1}^M$ from the noisy distribution $\mathbf{P}_{\rm noisy} = \mathbf{\Lambda}\mathbf{P}_{\rm ideal}$.
We then learn the MPS model $\mathbf{P}_{\rm MPS}$ for the ideal distribution $\mathbf{P}_{\rm ideal}$ by minimizing the negative log-likelihood function $\mathbb{L}$ in Eq.~\eqref{Equ: Loss}, and generate $M^{\prime}=M=50000$ new samples $\{\mathbf{y}^{(m)}_{\rm MPS}\}_{m=1}^{M^{\prime}}$ from the learned MPS model $\mathbf{P}_{\rm MPS}$.
The results in Fig.~\ref{Fig: RCS}(a) show that the XEB fidelity of the noisy samples is improved by the REM method for large $D$, where the crossing between the noisy and REM results occurs at a bond dimension $D\sim 10$ for $N=10$ qubits.
The scaling of the crossing bond dimension with system size $N$ is further shown in Fig.~\ref{Fig: RCS}(b), where the crossing bond dimension, although smaller than that required to accurately represent and simulate random circuits ($D=2^{N/2}$), still grows exponentially with $N$.

These results demonstrate that the improvement of the XEB fidelity by the REM method is limited by the expressibility of the MPS model for the ideal distribution $\mathbf{P}_{\rm ideal}$.
Nevertheless, as we have mentioned above, not all applications of sampling problems require REM.
The specific case we consider here is to use the XEB fidelity to demonstrate the supremacy of quantum computers, which does not have practical applications itself.
Therefore, instead of generating new samples that are closer to ideal distribution, we can use the noisy samples to infer the XEB fidelity in the case without readout errors.
This can be achieved by a completely classical postprocessing of the noisy samples $\{\mathbf{x}^{(m)}_{\rm noisy}\}_{m=1}^M$ and the learned MPO readout error model $\mathbf{\Lambda}_{\rm MPO}$, without explicitly learning the ideal distribution $\mathbf{P}_{\rm ideal}$.
Specifically, since XEB fidelity is a linear function of samples, we can directly apply the formula in Eq.~\eqref{Equ: nonlocal} to calculate the mitigated XEB fidelity as
\begin{align}
    F_{\rm XEB, REM} = \frac{2^N}{M}\sum_{m=1}^M \sum_{\mathbf{y}} \mathbf{P}_{\rm ideal}(\mathbf{y}) \left(\mathbf{\Lambda}_{\rm MPO}^{-1}\right)_{\mathbf{y}, \mathbf{x}^{(m)}_{\rm noisy}} - 1,
\end{align}
where the matrix multiplication can be efficiently calculated by contracting the corresponding tensor networks.
The results in Fig.~\ref{Fig: RCS}(c) show that the mitigated XEB fidelity is significantly improved compared to the noisy one, and is close to the readout-error-free value (which is itself not exactly one at this finite size), demonstrating the effectiveness of our method for eliminating the readout effects in the evaluation of XEB fidelity in random circuit sampling.

\section{REM under random measurements}
In this section, we discuss the REM for random measurement problems, where the measurement basis is randomly chosen for each shot.
This problem is of great importance in various applications, such as the classical shadow estimation~\cite{Huang2020} and the randomized measurement protocols for quantum state tomography~\cite{Torlai2023}.
Different from the previous applications where the measurement basis is fixed, the random measurement problems make the REM more challenging since each noisy sample is generated from an individual measurement basis, from which one cannot learn the ideal distribution under this concrete basis or obtain samples from it.
Therefore, instead of proposing a unified framework for REM under random measurements, we will discuss two specific applications, namely classical shadow estimation and quantum state tomography.
In these tasks, we try to obtain the ideal estimation of target quantities, such as the expectation values of observables in the classical shadow estimation or the reconstructed density matrix in the quantum state tomography, without explicitly learning the ideal distribution under each measurement basis.

\subsection{REM for classical shadow estimation}
Here, we focus on the classical shadow estimation for the expectation value of a set of observables $\braket{O}$.
We first review the classical shadow estimation method without REM, and then discuss how to integrate the learned MPO readout error model into this method to mitigate the readout errors.

In the classical shadow estimation method, one first applies a random unitary transformation $U^{(m)}$ to the quantum state $\rho$ before the measurement, leading to a new state $U^{(m)}\rho U^{(m)\dagger}$.
Then, one performs a projective measurement in the computational basis, obtaining a sample $\mathbf{x}^{(m)}$ drawn from the distribution $\mathbf{P}^{(m)}(\mathbf{x}) = \braket{\mathbf{x}| U^{(m)}\rho U^{(m)\dagger}|\mathbf{x}}$.
Define the map
\begin{align}
\begin{aligned}
    \mathcal{M}(\rho) &= \mathbb{E}_{U} \sum_{\mathbf{x}} \braket{\mathbf{x}| U \rho U^{\dagger}|\mathbf{x}} U^{\dagger} \ket{\mathbf{x}}\!\bra{\mathbf{x}} U\\
    &= \mathbb{E}_{U} \sum_{\mathbf{x}} \mathbf{P}_{U}(\mathbf{x}) U^{\dagger} \ket{\mathbf{x}}\!\bra{\mathbf{x}} U,\label{Equ: Map}
\end{aligned}
\end{align}
where $\mathbb{E}_U$ denotes the statistical expectation under random unitaries $U$.
Notably, $\mathcal{M}$ is a linear map on the operator space that is determined solely by the ensemble of random unitaries $\{U\}$, and is completely independent of the unknown state $\rho$.
For the commonly used random-unitary ensembles, both $\mathcal{M}$ and its inverse $\mathcal{M}^{-1}$ admit analytical forms derived from the properties of unitary designs~\cite{Huang2020}, as we show explicitly below for the case of random Pauli measurements.

The reconstruction of $\rho$ from the experiment then combines two independent ingredients.
On the one hand, the analytical form of the inverse map $\mathcal{M}^{-1}$ is fixed once the random-unitary ensemble is chosen, and can therefore be regarded as known prior to the experiment, without any reference to the measurement data.
On the other hand, the operator on the right-hand side of Eq.~\eqref{Equ: Map} for the specific unknown state $\rho$ is estimated purely from the experimental snapshots as
\begin{align}
    \mathcal{M}(\rho) \approx \frac{1}{M}\sum_{m=1}^M U^{(m)\dagger} \ket{\mathbf{x}^{(m)}}\!\bra{\mathbf{x}^{(m)}} U^{(m)},
\end{align}
where both the expectation over $U$ and the sum over $\mathbf{x}$ are replaced by the average over the samples $\{\mathbf{x}^{(m)}, U^{(m)}\}_{m=1}^M$.
Combining these two ingredients, one applies the analytically known inverse map $\mathcal{M}^{-1}$ to the estimated $\mathcal{M}(\rho)$ to reconstruct the density matrix as
\begin{align}
    \rho \approx \frac{1}{M}\sum_{m=1}^M \mathcal{M}^{-1}\left(U^{(m)\dagger} \ket{\mathbf{x}^{(m)}}\!\bra{\mathbf{x}^{(m)}} U^{(m)}\right).\label{Equ: Original Shadow}
\end{align}
Finally, one obtains an unbiased estimation of the expectation value of any observable $O$ as $\braket{O} = \Tr(O \rho)$.

An experimentally friendly and thus widely used method for classical shadow estimation is based on random Pauli measurements.
In this approach, to measure in the Pauli basis $\bigotimes_{i=1}^N S^{(m)}_i$, where $S^{(m)}_i \in \{X, Y, Z\}$ is the Pauli operator for the $i$-th qubit in the $m$-th shot, one can apply a local unitary rotation to each qubit before the measurement
\begin{align}
    U^{(m)} = \bigotimes_{i=1}^N U^{(m)}_i = \bigotimes_{i=1}^N U_{S^{(m)}_i},
\end{align}
where each single-qubit unitary $U_{S^{(m)}_i}$ is used to rotate the measurement basis from $Z$ to any Pauli basis $S^{(m)}_i$.
A convenient choice is $U_Z=I$, $U_X=H$, and $U_Y=S^\dagger H$.
In this case, both the linear map $\mathcal{M}$ and its inverse $\mathcal{M}^{-1}$ are the tensor products of single-qubit maps, i.e.,
\begin{align}
    &\mathcal{M} = \bigotimes_{i=1}^N \mathcal{M}_i, \quad \mathcal{M}_i(\sigma_i) = \frac{\sigma_i + \Tr(\sigma_i)I}{3},\label{Equ: Pauli Map}\\
    &\mathcal{M}^{-1} = \bigotimes_{i=1}^N \mathcal{M}_i^{-1}, \quad \mathcal{M}_i^{-1}(\sigma_i) = 3\sigma_i - \Tr(\sigma_i)I,
\end{align}
where $\sigma_i$ is a single-qubit operator defined on the $i$-th qubit.
In this case, the classical shadow samples $\{\mathbf{x}^{(m)}, U^{(m)}\}_{m=1}^M$ can be used to reconstruct the density matrix $\rho$ as
\begin{align}
    \rho \approx \frac{1}{M}\sum_{m=1}^M \bigotimes_{i=1}^N \left(3U^{(m)\dagger}_i \ket{x^{(m)}_i}\!\bra{x^{(m)}_i} U^{(m)}_i-I\right).
\end{align}

\begin{figure}
    \centering
    \includegraphics[width=\linewidth]{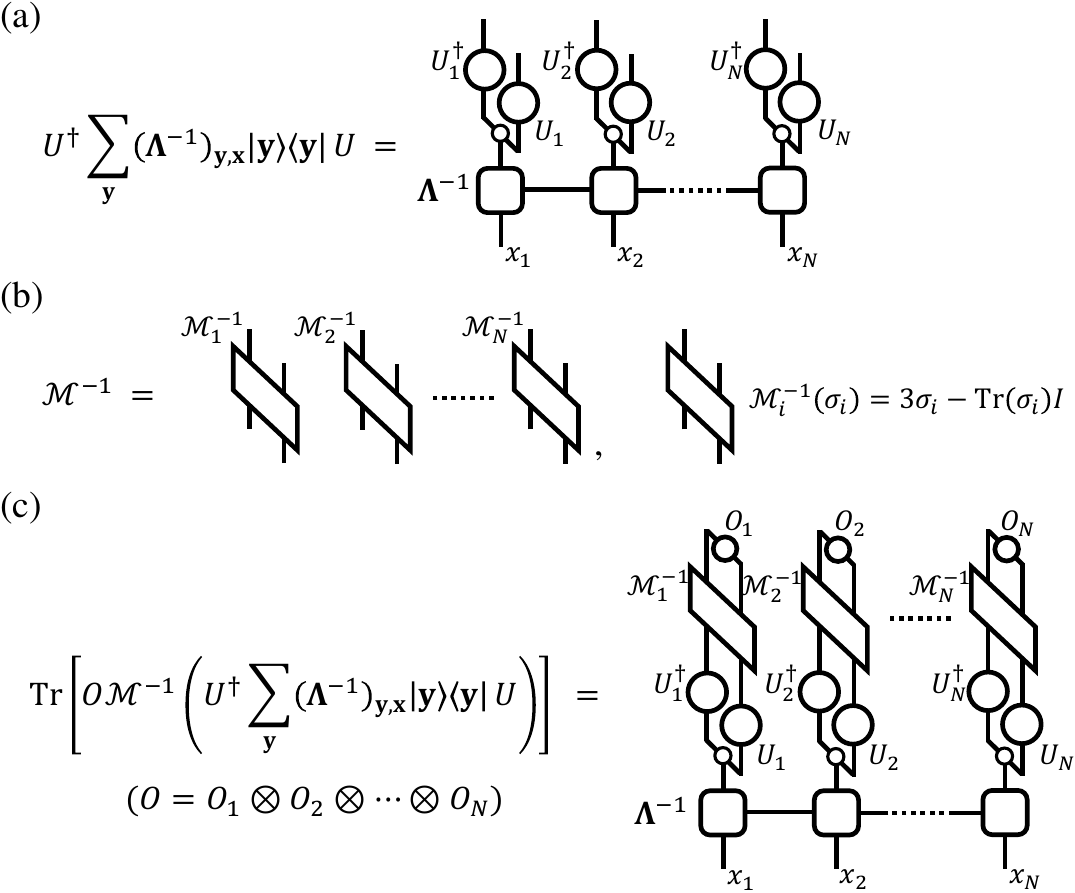}
    \caption{REM for classical shadow estimation.
    (a) The operator $U^{\dagger} \left[\sum_{\mathbf{y}} \left(\mathbf{\Lambda}^{-1}\right)_{\mathbf{y}, \mathbf{x}}  \ket{\mathbf{y}}\!\bra{\mathbf{y}}\right] U$ is spanned by the single-layered MPO $\mathbf{\Lambda}_{\rm MPO}^{-1}$.
    (b) $\mathcal{M}^{-1}$ represents a linear transformation on the operator space, which has a tensor product structure.
    (c) The expectation value of any observable $O$ can be estimated by contracting the corresponding tensor networks $\Tr\left[O \mathcal{M}^{-1}\left(\sum_{\mathbf{y}} \left(\mathbf{\Lambda}^{-1}\right)_{\mathbf{y}, \mathbf{x}} U^{\dagger} \ket{\mathbf{y}}\!\bra{\mathbf{y}} U\right)\right]$ for all samples $\{\mathbf{x}^{(m)}_{\rm noisy}, U^{(m)}\}_{m=1}^M$.}
    \label{Fig: Shadow}
\end{figure}
Now, we discuss how to mitigate the readout errors in the classical shadow estimation method.
We rewrite Eq.~\eqref{Equ: Map} as
\begin{align}
\begin{aligned}
    \mathcal{M}(\rho) &= \mathbb{E}_{U} \sum_{\mathbf{y}} \mathbf{P}_{U, \rm ideal}(\mathbf{y}) U^{\dagger} \ket{\mathbf{y}}\!\bra{\mathbf{y}} U\\
    &= \mathbb{E}_{U} \sum_{\mathbf{x}, \mathbf{y}} \left(\mathbf{\Lambda}^{-1}\right)_{\mathbf{y}, \mathbf{x}} \mathbf{P}_{U, \rm noisy}(\mathbf{x}) U^{\dagger} \ket{\mathbf{y}}\!\bra{\mathbf{y}} U\\
    &\approx \frac{1}{M} \sum_{m=1}^M \sum_{\mathbf{y}} \left(\mathbf{\Lambda}^{-1}\right)_{\mathbf{y}, \mathbf{x}^{(m)}_{\rm noisy}} U^{(m)\dagger} \ket{\mathbf{y}}\!\bra{\mathbf{y}} U^{(m)},
\end{aligned}
\end{align}
from which the expectation value of any observable $O$ can be estimated as
\begin{align}
\begin{aligned}
    \braket{O} & = \Tr\left(O \rho\right)\\
    & \approx \frac{1}{M} \sum_{m=1}^M \Tr\left[O \mathcal{M}^{-1}\left(\sum_{\mathbf{y}} \left(\mathbf{\Lambda}^{-1}\right)_{\mathbf{y}, \mathbf{x}^{(m)}_{\rm noisy}} U^{(m)\dagger} \ket{\mathbf{y}}\!\bra{\mathbf{y}} U^{(m)}\right)\right].
    \label{Equ: Shadow QEM}
\end{aligned}
\end{align}
This formula provides a general framework for REM in the classical shadow estimation method.

The main challenge lies in the efficient calculation of Eq.~\eqref{Equ: Shadow QEM} using the inverse of the MPO readout error model $\mathbf{\Lambda}_{\rm MPO}^{-1}$.
As shown in Fig.~\ref{Fig: Shadow}(a), the inverse MPO $\mathbf{\Lambda}_{\rm MPO}^{-1}$ is used to generate the operator $\sum_{\mathbf{y}} \left(\mathbf{\Lambda}^{-1}_{\rm MPO}\right)_{\mathbf{y}, \mathbf{x}^{(m)}_{\rm noisy}} \ket{\mathbf{y}}\!\bra{\mathbf{y}}$ for a given sample $\mathbf{x}^{(m)}_{\rm noisy}$.
This operator is then contracted with the unitary transformation $U^{(m)}$ and the inverse map $\mathcal{M}^{-1}$ shown in Fig.~\ref{Fig: Shadow}(b) to obtain the contribution of this sample to the estimation of $\rho$.
Finally, the expectation value of the observable $O$ can be estimated by averaging over the corresponding tensor networks for all samples $\{\mathbf{x}^{(m)}_{\rm noisy}, U^{(m)}\}_{m=1}^M$.

\subsection{REM for quantum state tomography}
\begin{figure*}
    \centering
    \includegraphics[width=\linewidth]{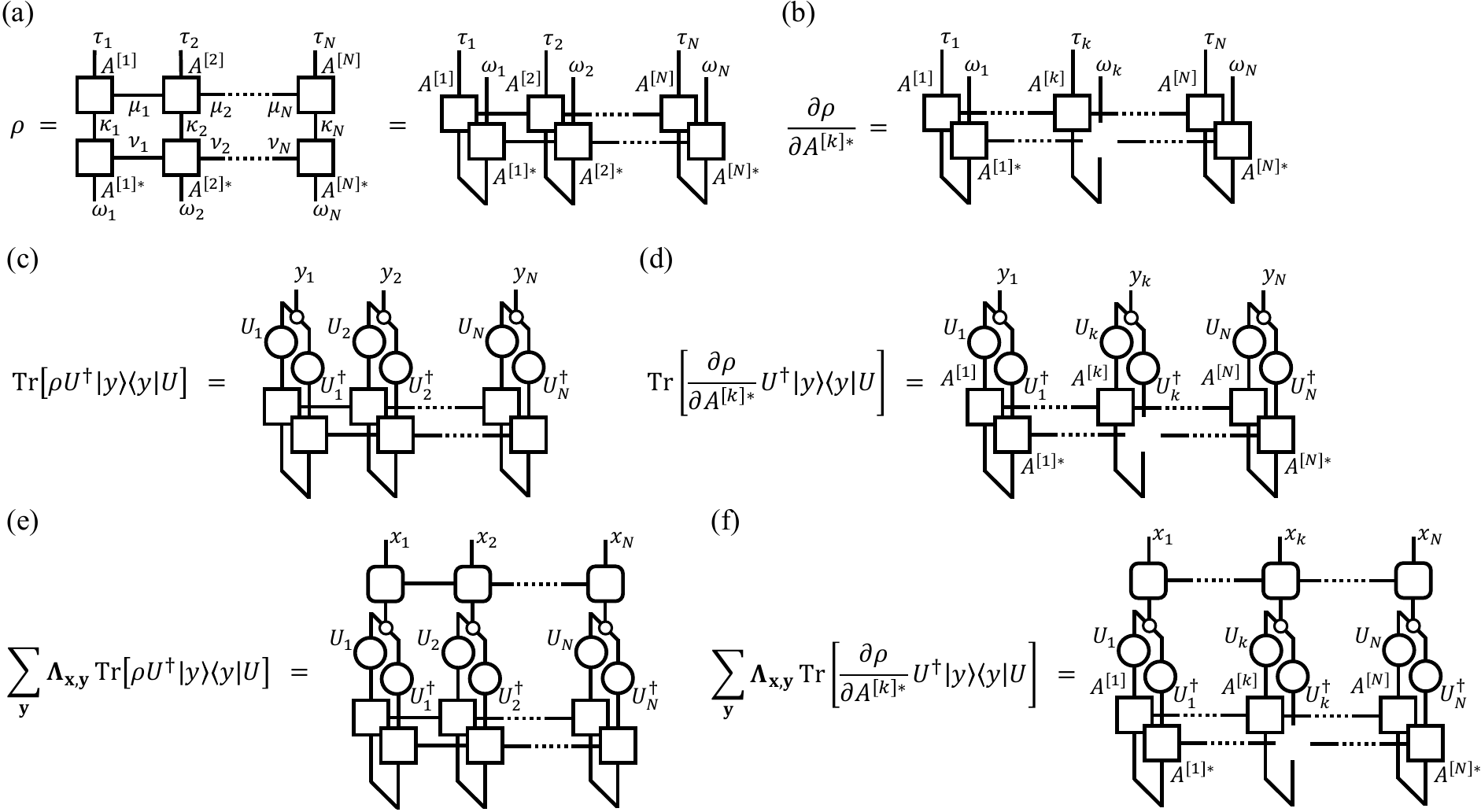}
    \caption{REM for quantum state tomography.
    (a) The LPDO represents a parameterized ansatz for the density matrix $\rho$.
    (b) The tensor network for the gradient $\partial \rho/\partial A^{[k]*}$.
    (c) $\Tr\left(\rho   U^{\dagger} \ket{\mathbf{y}}\!\bra{\mathbf{y}} U\right)$ corresponds to a vector regarding the index $\mathbf{y}$.
    (d) The gradient $\Tr\left(\partial \rho/\partial A^{[k]*}   U^{\dagger} \ket{\mathbf{y}}\!\bra{\mathbf{y}} U\right)$ can be efficiently calculated by contracting the corresponding tensor networks.
    (e) The summation over $\mathbf{y}$ in $ \sum_{\mathbf{y}} \mathbf{\Lambda}_{\mathbf{x}, \mathbf{y}}\Tr\left(\rho   U^{\dagger} \ket{\mathbf{y}}\!\bra{\mathbf{y}} U\right)$ can be efficiently calculated by contracting the tensor networks shown in (c) and the readout error matrix $\mathbf{\Lambda}_{\rm MPO}$.
    (f) The gradient $ \sum_{\mathbf{y}} \mathbf{\Lambda}_{\mathbf{x}, \mathbf{y}}\Tr\left(\partial \rho/\partial A^{[k]*} U^{\dagger} \ket{\mathbf{y}}\!\bra{\mathbf{y}} U\right)$ can be efficiently calculated by contracting the tensor networks shown in (d) and $\mathbf{\Lambda}_{\rm MPO}$.}
    \label{Fig: Tomography}
\end{figure*}

Quantum state tomography is the task of reconstructing the density matrix $\rho$ of a quantum state from measurement data, where the unknown state is modeled by a parameterized ansatz $\rho(\theta)$, such as tensor network and neural network models.
Some approaches utilize local reduced density matrices to reconstruct the global state, while others optimize the parameters $\theta$ to learn the measurement statistics.
To account for the quantumness of the state, the latter approaches typically involve random measurements, making the REM more challenging.

Here, we focus on the learning-based quantum state tomography method proposed in Ref.~\cite{Torlai2023}, where the density matrix $\rho$ is modeled by a locally purified density operator (LPDO)~\cite{Verstraete2004}.
The LPDO is a tensor network ansatz for mixed states, reading as
\begin{align}
\begin{aligned}
    \rho = &\,\sum_{\{\bm{\tau}, \bm{\omega}\}} \sum_{\{\bm{\mu}, \bm{\nu}, \bm{\kappa}\}} \prod_{k=1}^N\left(A^{[k]}\right)^{\mu_{k}, \mu_{k+1}}_{\tau_k, \kappa_k}\left(A^{[k]*}\right)^{\nu_{k}, \nu_{k+1}}_{\omega_k, \kappa_k}\\
    &\,\ket{\tau_1, \cdots, \tau_N}\!\bra{\omega_1, \cdots, \omega_N},
\end{aligned}
\end{align}
where a Kraus index $\kappa_k$ is introduced to each site to represent the ancillary degree of freedom, as shown in Fig.~\ref{Fig: Tomography}(a).
This ansatz has wide applications in quantum state tomography~\cite{Torlai2023, Guo2024a}, simulation of dissipative systems~\cite{Verstraete2004, Werner2016, Guo2024}, and topological phases in open quantum systems~\cite{Guo2025}.

To learn the optimal parameters of the LPDO model $\{A^{[k]}\}$, one performs random measurements on the quantum state $\rho$ to obtain a set of samples $\{\mathbf{x}^{(m)}, U^{(m)}\}_{m=1}^M$, where $U^{(m)}$ is the random unitary transformation to rotate the measurement basis similar to the classical shadow estimation method.
Next, one can calculate the probability of generating the sample $\mathbf{x}^{(m)}$ under the measurement basis defined by $U^{(m)}$ as
\begin{align}
    \mathbf{P}_{U^{(m)}}(\mathbf{x}^{(m)}) = \frac{\Tr\left(\rho U^{(m)\dagger} \ket{\mathbf{x}^{(m)}}\!\bra{\mathbf{x}^{(m)}} U^{(m)} \right)}{\Tr(\rho)},
\end{align}
as shown in Fig.~\ref{Fig: Tomography}(c).
Finally, the parameters $\{A^{[k]}\}$ can be optimized by minimizing the negative log-likelihood function
\begin{align}
    \mathbb{L} = -\frac{1}{M}\sum_{m=1}^M \log \mathbf{P}_{U^{(m)}}(\mathbf{x}^{(m)})
    \label{Equ: Tomography Loss}
\end{align}
with gradient-based optimization methods, such as the Adam optimizer.
Notably, since we are working with complex numbers, the local tensor is updated by
\begin{align}
    A^{[k]} \rightarrow A^{[k]} - \eta \frac{\partial \mathbb{L}}{\partial A^{[k]*}},
\end{align}
where $\eta$ is the learning rate and $\partial \mathbb{L}/\partial A^{[k]*}$ is the gradient of the loss function with respect to the complex conjugate of the local tensor $A^{[k]}$.
The gradient of the loss function is expressed as
\begin{align}
\begin{aligned}
    \frac{\partial \mathbb{L}}{\partial A^{[k]*}} & = -\frac{1}{M}\sum_{m=1}^M \frac{\Tr\left(\frac{\partial \rho}{\partial A^{[k]*}} U^{(m)\dagger} \ket{\mathbf{x}^{(m)}}\!\bra{\mathbf{x}^{(m)}} U^{(m)}\right)}{\Tr\left(\rho U^{(m)\dagger} \ket{\mathbf{x}^{(m)}}\!\bra{\mathbf{x}^{(m)}} U^{(m)} \right)}\\
    & + \frac{1}{\Tr(\rho)} \Tr\left(\frac{\partial \rho}{\partial A^{[k]*}}\right),
\end{aligned}
\end{align}
which can be efficiently calculated by contracting the corresponding tensor networks in Fig.~\ref{Fig: Tomography}(b-d).

Now, suppose the measurement data $\{\mathbf{x}^{(m)}_{\rm noisy}, U^{(m)}\}_{m=1}^M$ are affected by readout errors, where each sample $\mathbf{x}^{(m)}_{\rm noisy}$ is drawn from the noisy distribution $\mathbf{P}_{U^{(m)}, \rm noisy} = \mathbf{\Lambda}\mathbf{P}_{U^{(m)}, \rm ideal}$.
Instead of mitigating the readout errors in the measurement data $\{\mathbf{x}^{(m)}_{\rm noisy}, U^{(m)}\}_{m=1}^M$ by, e.g., applying the inverse of the readout error model $\mathbf{\Lambda}^{-1}$ to each sample, we can directly integrate $\mathbf{\Lambda}$ into the optimization of the LPDO model.
Specifically, we can rewrite the probability of generating the sample $\mathbf{x}^{(m)}_{\rm noisy}$ under the measurement basis defined by $U^{(m)}$ as
\begin{align}
\begin{aligned}
    \mathbf{P}_{U^{(m)}, \rm noisy}(\mathbf{x}^{(m)}_{\rm noisy}) &= \sum_{\mathbf{y}} \mathbf{\Lambda}_{\mathbf{x}^{(m)}_{\rm noisy}, \mathbf{y}} \mathbf{P}_{U^{(m)}, \rm ideal}(\mathbf{y})\\
    & =  \sum_{\mathbf{y}} \mathbf{\Lambda}_{\mathbf{x}^{(m)}_{\rm noisy}, \mathbf{y}}\frac{\Tr\left(\rho   U^{(m)\dagger} \ket{\mathbf{y}}\!\bra{\mathbf{y}} U^{(m)}\right)}{\Tr(\rho)},
\end{aligned}
\end{align}
as shown in Fig.~\ref{Fig: Tomography}(e).

Therefore, the negative log-likelihood function can be rewritten as
\begin{align}
    \mathbb{L} = -\frac{1}{M}\sum_{m=1}^M \log \left[\sum_{\mathbf{y}} \mathbf{\Lambda}_{\mathbf{x}^{(m)}_{\rm noisy}, \mathbf{y}}\frac{\Tr\left(\rho   U^{(m)\dagger} \ket{\mathbf{y}}\!\bra{\mathbf{y}} U^{(m)}\right)}{\Tr(\rho)}\right],
    \label{Equ: Tomography Loss QEM}
\end{align}
whose gradient with respect to the local tensor $A^{[k]*}$ reads as
\begin{align}
\begin{aligned}
    \frac{\partial \mathbb{L}}{\partial A^{[k]*}} & = -\frac{1}{M}\sum_{m=1}^M \frac{\sum_{\mathbf{y}} \mathbf{\Lambda}_{\mathbf{x}^{(m)}_{\rm noisy}, \mathbf{y}} \Tr\left(\frac{\partial \rho}{\partial A^{[k]*}}   U^{(m)\dagger} \ket{\mathbf{y}}\!\bra{\mathbf{y}} U^{(m)}\right)}{\sum_{\mathbf{y}} \mathbf{\Lambda}_{\mathbf{x}^{(m)}_{\rm noisy}, \mathbf{y}}\Tr\left(\rho   U^{(m)\dagger} \ket{\mathbf{y}}\!\bra{\mathbf{y}} U^{(m)}\right)}\\
    & + \frac{1}{\Tr(\rho)}\Tr\left(\frac{\partial \rho}{\partial A^{[k]*}}\right).
\end{aligned}
\end{align}
This gradient can be efficiently calculated by contracting the corresponding tensor networks shown in Fig.~\ref{Fig: Tomography}(b, e, f).
Finally, we can optimize the parameters of the LPDO model by minimizing this modified loss function using the Adam optimizer with the same hyperparameters as before.

\subsection{Numerical simulations}
\begin{figure*}
    \centering
    \includegraphics*[width=0.66\linewidth]{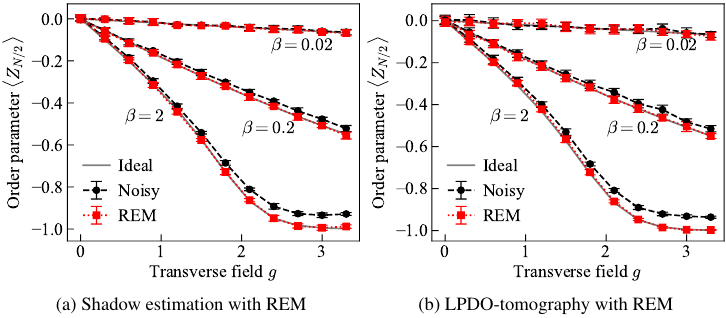}
    \caption{Numerical simulations for REM under random measurements.
    (a) The estimation of the magnetization $\braket{Z_{N/2}}$ from the classical shadow samples with and without REM.
    (b) The estimation of the magnetization $\braket{Z_{N/2}}$ from the quantum state tomography method with and without REM.
    In both methods, the ideal value is calculated from the LPDO representation of the Gibbs state $\rho$ with $N=20$ and $D=32$.}
    \label{Fig: Random}
\end{figure*}
We perform numerical simulations for REM in classical shadow estimation and quantum state tomography.
In both tasks, we consider a finite-temperature Gibbs state $\rho = e^{-\beta H}/\Tr(e^{-\beta H})$ of a 1D XY model with transverse field
\begin{align}
    H = \sum_{i=1}^{N-1} \left(X_i X_{i+1} + Y_i Y_{i+1}\right) + g \sum_{i=1}^N Z_i,
\end{align}
under the open boundary condition.
To perform numerical simulations, we first implement imaginary-time evolution
\begin{align}
    e^{-\beta H} = \left(e^{-\Delta \tau H}\right)^{N_\tau} I \left(e^{-\Delta \tau H}\right)^{N_\tau}, \quad N_\tau = \frac{\beta}{2\Delta \tau},
\end{align}
to obtain the LPDO representation of the Gibbs state $\rho$ with $N=20$ and bond dimension $D=32$ using the second-order Trotter decomposition and $\Delta \tau = 0.01$~\cite{Verstraete2004}.
We then focus on the estimation of the magnetization $\braket{Z_{N/2}}$ of the central qubit in the following numerical simulations.

For the classical shadow estimation, we first generate $M=50000$ samples $\{\mathbf{y}^{(m)}, U^{(m)}\}_{m=1}^M$ from the Gibbs state $\rho$, and then apply the same readout error model as used in Fig.~\ref{Fig: Sim1}(a) to these samples to obtain the noisy samples $\{\mathbf{x}^{(m)}_{\rm noisy}, U^{(m)}\}_{m=1}^M$.
The MPO readout error model $\mathbf{\Lambda}_{\rm MPO}$ with $\chi=4$ is also learned by $M=50000$ samples from calibration experiments.
Finally, we estimate the magnetization $\braket{Z_{N/2}}$ using the original classical shadow estimation method without REM in Eq.~\eqref{Equ: Original Shadow} as well as the modified method with REM in Eq.~\eqref{Equ: Shadow QEM}, where the expectation value $\braket{Z_{N/2}}$ is estimated by contracting the corresponding tensor networks for all samples $\{\mathbf{x}^{(m)}_{\rm noisy}, U^{(m)}\}_{m=1}^M$ shown in Fig.~\ref{Fig: Shadow}(c).
These two results are compared with the ideal value of $\braket{Z_{N/2}}$ directly calculated from the LPDO representation of the Gibbs state $\rho$ in Fig.~\ref{Fig: Random}(a) to evaluate the performance of the REM method.
Notably, due to the flipping symmetry between $0$ and $1$ in the readout error model, the estimation of $\braket{Z_{N/2}}$ without REM is biased towards zero, and this bias is much larger for the low-temperature regime with a larger magnetization.
Nevertheless, the REM method can significantly improve the estimation of $\braket{Z_{N/2}}$ compared to the original method without REM across different parameter regimes, demonstrating the effectiveness of our method.

For the quantum state tomography, we use the same set of noisy samples $\{\mathbf{x}^{(m)}_{\rm noisy}, U^{(m)}\}_{m=1}^M$ to optimize the parameters of the LPDO model by minimizing the original loss function in Eq.~\eqref{Equ: Tomography Loss} and the modified loss function with REM in Eq.~\eqref{Equ: Tomography Loss QEM}.
The estimation of $\braket{Z_{N/2}}$ from the optimized LPDO model with and without REM is compared with the ideal value calculated from the LPDO representation of the Gibbs state $\rho$ in Fig.~\ref{Fig: Random}(b).
The results show that the REM method can significantly improve the estimation of $\braket{Z_{N/2}}$ compared to the original method without REM, demonstrating the effectiveness of our method in quantum state tomography.
Notably, compared to the above shadow estimation method, the uncertainty of the estimation from the tomography method is slightly larger, which is due to the fact that the tomography method tries to learn a global model for the density matrix $\rho$ while the shadow estimation method directly estimates the expectation value of the target observable and is more suitable for predicting local observables~\cite{Huang2020}.

\section{REM for QEC decoding}
In general, implementing various error mitigation techniques on logical qubits is a common goal, which can significantly improve the performance of logical qubits and thus reduce the resource overhead for QEC~\cite{Suzuki2022, Wahl2023}.
In this section, we discuss a much deeper integration of REM with the decoding process in QEC, where the learned readout error model is directly fused into the maximum-likelihood (ML) decoding objective to mitigate the readout errors in the syndrome data.
This approach is particularly well matched to tensor-network decoders~\cite{Bravyi2014, Ferris2014, Chubb2021}, where both the code prior and the readout model can be represented and contracted as tensor networks.

\begin{figure*}[t]
    \centering
    \includegraphics[width=\linewidth]{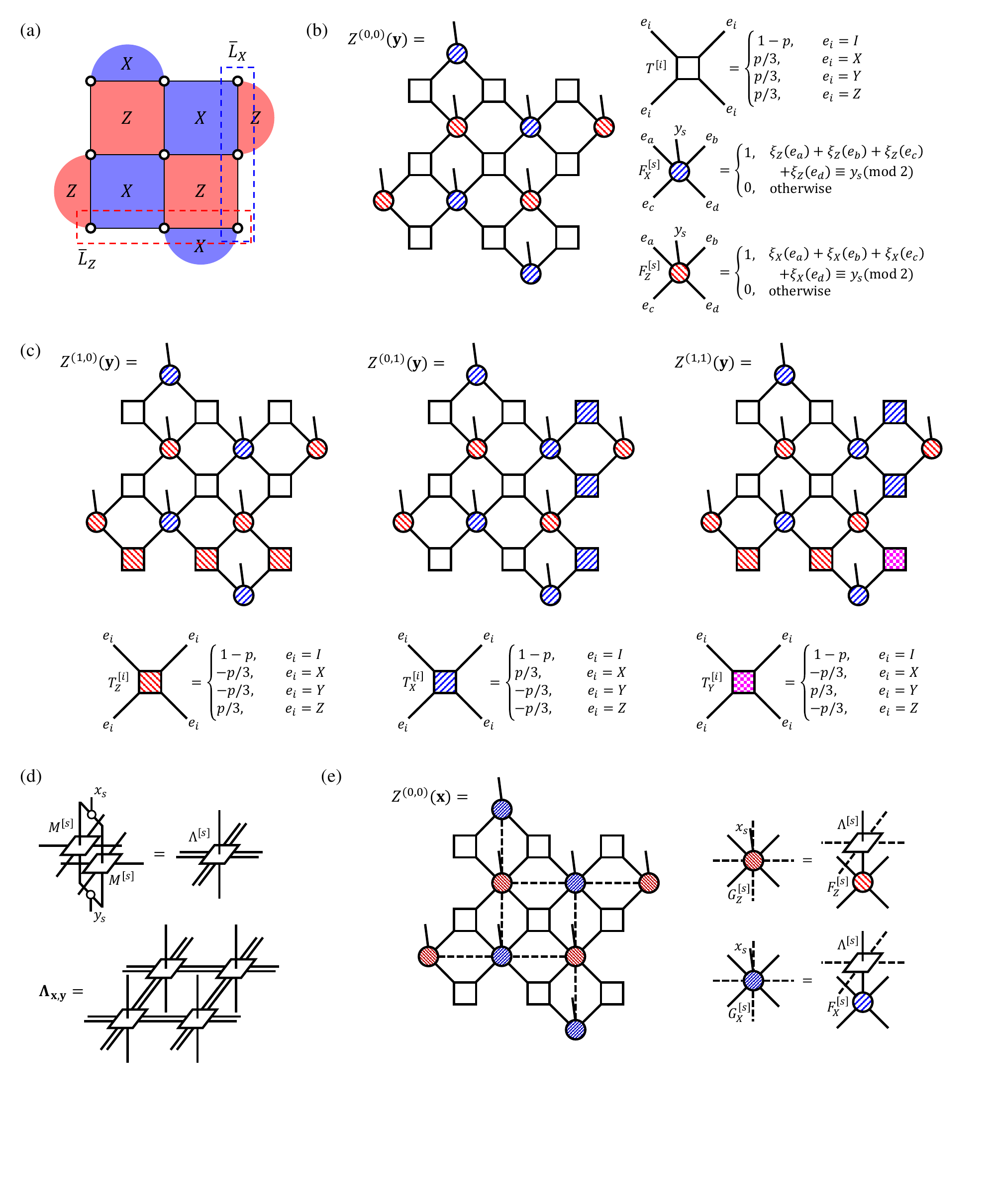}
    \caption{Schematic for QEC decoding with correlated readout mitigation, illustrated on a $d=3$ rotated surface code.
    (a) Data qubits and $X/Z$ stabilizers.
    Data qubits that support the logical operators $\bar L_Z$ and $\bar L_X$ are highlighted in red and blue dashed rectangles, respectively.
    (b) Tensor-network representation of the partition function $Z^{(0,0)}(\mathbf{y})$.
    Each data-qubit tensor $T^{[i]}$ encodes $P(e_i)$ over $e_i\in\{I,X,Y,Z\}$.
    Each $Z$-type ($X$-type) syndrome tensor $F_Z^{[s]}$ ($F_X^{[s]}$) imposes the parity constraint on $\xi_X(e)$ ($\xi_Z(e)$) values of its four neighboring data qubits to detect $X$ ($Z$) physical errors.
    (c) Tensor-network representation of the other three signed partition functions $Z^{(a,b)}(\mathbf{y})$ for $a,b\in\{0,1\}$.
    $T_Z^{[i]}$, $T_X^{[i]}$, and $T_Y^{[i]}$ implement the sign insertions $(-1)^{\xi_X(e_i)}$, $(-1)^{\xi_Z(e_i)}$, and $(-1)^{\xi_X(e_i)+\xi_Z(e_i)}$ on $\mathrm{supp}(\bar L_Z)$, $\mathrm{supp}(\bar L_X)$, and their intersection to detect $X$, $Z$, and $Y$ logical errors, respectively.
    (d) PEPO representation of the correlated readout error model $\mathbf{\Lambda}_{\mathbf{x},\mathbf{y}}$.
    (e) Fusion of the PEPO readout model with the decoder PEPS.}
    \label{Fig: QEC Sketch}
\end{figure*}

\subsection{Surface code}
We focus on the rotated surface code of distance $d$~\cite{Fowler2012, Google2023} [Fig.~\ref{Fig: QEC Sketch}(a) for $d=3$], where $N=d^2$ data qubits are placed on the vertices of a $d\times d$ square lattice.
The code is defined by $N-1$ stabilizer generators arranged in a checkerboard pattern on the plaquettes of this lattice.
Each $X$/$Z$-type stabilizer is the product of Pauli-$X$/$Z$ operators acting on the data qubits around a plaquette, where a stabilizer in the bulk (on the boundary) acts on four (two) data qubits, as illustrated in Fig.~\ref{Fig: QEC Sketch}(a).
The code space is defined as the simultaneous $+1$ eigenspace of all stabilizers.
Since the $N$ physical qubits are subject to $N-1$ independent stabilizer constraints, exactly one logical qubit is encoded, with logical operators $\bar L_X$ and $\bar L_Z$ realized as string-like products of single-qubit $X$'s and $Z$'s connecting two opposite boundaries of the lattice.
In experiments, each stabilizer is measured with the help of a syndrome qubit located at the center of the corresponding plaquette (i.e., on the dual lattice) using a stabilizer measurement circuit, and the measured stabilizer values are collected to detect the errors on the data qubits.

For concreteness, we consider an independent single-qubit depolarizing noise on each data qubit, generating an error pattern $\mathbf{e}\in\{I,X,Y,Z\}^N$ with prior probability
\begin{align}
    P(\mathbf{e}) = \prod_{i=1}^{N} P(e_i),\qquad P(e_i)=\begin{cases}1-p & e_i=I\\ p/3 & e_i\in\{X,Y,Z\}\end{cases},
\end{align}
where $p$ is the physical error rate.
We use the symplectic decomposition of single-qubit Pauli errors, i.e., every single-qubit Pauli operator can be written (up to an irrelevant global phase) as $e = X^{\xi_X(e)} Z^{\xi_Z(e)}$, where the two binary indicators $\xi_X(e),\xi_Z(e)\in\{0,1\}$ represent whether $e$ contains an $X$ or $Z$ component.
In particular, since $Y\propto XZ$ contains both an $X$ and a $Z$ component, these indicators read $\xi_X(I,X,Y,Z)=(0,1,1,0)$ and $\xi_Z(I,X,Y,Z)=(0,0,1,1)$.

These errors are detected by the stabilizer measurements through commutation relations, i.e., a stabilizer outcome is flipped iff the error pattern $\mathbf{e}$ anticommutes with this stabilizer.
Since anticommutation only arises between Pauli-$X$ and Pauli-$Z$ acting on the same qubit, an $X$-type stabilizer flips its outcome whenever the number of $Z$-components of the error [$e_i$ with $\xi_Z(e_i)=1$] on its support is odd, and a $Z$-type stabilizer flips on the parity of $X$-components.
For instance, a single $Y$ error on a bulk data qubit that carries both components flips the outcomes of all four surrounding stabilizers.
Therefore, each data error configuration $\mathbf{e}$ generates a definite syndrome (combining both $X$- and $Z$-stabilizer outcomes) denoted by $\boldsymbol\alpha(\mathbf{e})$.

On the other hand, the syndrome does not uniquely determine the error pattern.
Two error patterns related by a product of stabilizers generate the same syndrome and act identically on the code space, and are thus physically equivalent.
Consequently, all error patterns consistent with a given syndrome split into four equivalence classes $(c_X,c_Z)\in\{0,1\}^2$, distinguished by whether the error additionally implements a logical-$X$ and/or logical-$Z$ operation on the encoded qubit.
To determine the logical class that an error pattern $\mathbf{e}$ belongs to, we use its commutation relations with the logical operators.
Since $\mathbf{e}$ contains a logical-$X$ error iff it anticommutes with $\bar L_Z$, and similarly for logical-$Z$ via $\bar L_X$, the class indicators are
\begin{align}
    c_X(\mathbf{e}) = \bigoplus_{i\in\mathrm{supp}(\bar L_Z)} \xi_X(e_i),\qquad
    c_Z(\mathbf{e}) = \bigoplus_{i\in\mathrm{supp}(\bar L_X)} \xi_Z(e_i),
\end{align}
where $\mathrm{supp}(\bar L)$ denotes the support of the operator $\bar L$, i.e., the set of data qubits on which it acts nontrivially, and $\oplus$ denotes addition modulo $2$.
Intuitively, each qubit $i\in\mathrm{supp}(\bar L_Z)$ on which the error carries an $X$-component contributes one anticommuting factor with the corresponding $Z$ operator in $\bar L_Z$, so $\mathbf{e}$ anticommutes with $\bar L_Z$ iff the total parity $c_X(\mathbf{e})$ is odd (and analogously for $c_Z$).

\subsection{Tensor-network maximum-likelihood decoding}
Let $\mathbf{y}$ be the observed syndrome.
Without readout errors, maximum-likelihood (ML) decoding amounts to computing the four partition functions~\cite{Bravyi2014,Ferris2014,Chubb2021}
\begin{align}
    Z_{c_X,c_Z}(\mathbf{y}) = \sum_{\mathbf{e}\in (c_X,c_Z)} P(\mathbf{e})\,\delta\!\left[\boldsymbol\alpha(\mathbf{e}),\mathbf{y}\right],
    \label{Equ: QEC Partition}
\end{align}
where the sum runs over all error patterns $\mathbf{e}$ belonging to the logical class $(c_X,c_Z)$, and the Kronecker delta $\delta\left[\boldsymbol\alpha(\mathbf{e}),\mathbf{y}\right]$ further restricts the sum to those consistent with the observed syndrome.
In other words, $Z_{c_X,c_Z}(\mathbf{y})$ is the total probability that an error from the class $(c_X,c_Z)$ occurs and produces the syndrome $\mathbf{y}$.
The recovery operation is then chosen from the class with the largest weight, which succeeds iff the actual error indeed belongs to this class.

The idea of recasting the ML decoding of surface codes into 2D tensor-network contractions was established in Refs.~\cite{Bravyi2014, Ferris2014, Chubb2021}, where the coset probability of each logical class is evaluated separately.
Here, we construct an equivalent tensor-network realization based on the signed partition functions $Z^{(a,b)}$, whose key feature is that the dependence on the syndrome $\mathbf{y}$ is completely localized on the syndrome tensors $F^{[s]}$ defined below.
This structure is essential for our purpose, as it allows the readout error model to be fused locally into the decoder, as discussed in the following subsections.
Specifically, instead of directly summing over each class as in Eq.~\eqref{Equ: QEC Partition}, it is more convenient to define the four signed partition functions for $a,b\in\{0,1\}$
\begin{align}
    Z^{(a,b)}(\mathbf{y}) = \sum_{\mathbf{e}} (-1)^{a\,c_X(\mathbf{e})+b\,c_Z(\mathbf{e})}\, P(\mathbf{e})\,\delta\!\left[\boldsymbol\alpha(\mathbf{e}),\mathbf{y}\right],
\end{align}
where the sum now runs over all error patterns without the restriction of logical classes.
Here, $a$ and $b$ are two auxiliary binary variables conjugate to the class labels $c_X$ and $c_Z$, controlling whether the sign factors $(-1)^{c_X(\mathbf{e})}$ and $(-1)^{c_Z(\mathbf{e})}$ are inserted into the unrestricted sum.
In other words, $\{Z^{(a,b)}\}$ is the discrete Fourier transform of $\{Z_{c_X,c_Z}\}$ over the group $\mathbb{Z}_2\times\mathbb{Z}_2$ of logical classes.
The advantage of this reformulation is that the unrestricted sum over $\mathbf{e}$ can be directly evaluated by contracting a tensor network composed of purely local tensors, as detailed below.
The four ML weights then follow from the inverse transform
\begin{align}
    Z_{c_X,c_Z}(\mathbf{y}) = \frac{1}{4}\sum_{a,b\in\{0,1\}}(-1)^{a c_X+b c_Z}\,Z^{(a,b)}(\mathbf{y}).
\end{align}
Therefore, it is sufficient to evaluate four signed partition functions to complete the decoding.

Specifically, $Z^{(0,0)}(\mathbf{y})$ can be expressed as the contraction of a projected entangled-pair state (PEPS) defined on the code lattice, where each data qubit corresponds to a local tensor $T^{[i]}$ and each syndrome qubit corresponds to a local tensor $F_X^{[s]}$ ($X$-stabilizer) or $F_Z^{[s]}$ ($Z$-stabilizer) projected onto the observed syndrome bit $y_s$, as shown in Fig.~\ref{Fig: QEC Sketch}(b).
The tensor $T^{[i]}$ sums over the local error variable $e_i\in\{I,X,Y,Z\}$ weighted by its prior probability $P(e_i)$, and copies $e_i$ itself onto its virtual indices shared with the neighboring syndrome tensors.
The syndrome tensor $F_Z^{[s]}$ then acts as a parity-constraint tensor, which extracts the $X$-components $\xi_X$ of the error variables on its neighboring data qubits and enforces their total parity to match the measured syndrome value $y_s$, and similarly for $F_X^{[s]}$ with the $Z$-components $\xi_Z$ of its neighboring error variables.
These two types of syndrome tensors are marked by the red ($Z$-stab) and blue ($X$-stab) circles in Fig.~\ref{Fig: QEC Sketch}(b).
Contracting all virtual indices of this tensor network automatically sums over all error configurations $\mathbf{e}$ consistent with the observed syndrome $\mathbf{y}$ and multiplies their prior probabilities $P(\mathbf{e})=\prod_i P(e_i)$, thereby exactly reproducing the definition of $Z^{(0,0)}(\mathbf{y})$.

The other three $Z^{(a,b)}$ are obtained by inserting the sign factor $(-1)^{a\,c_X(\mathbf{e})+b\,c_Z(\mathbf{e})}$, which factorizes into single-qubit signs on $\mathrm{supp}(\bar L_Z)$ and $\mathrm{supp}(\bar L_X)$ and is thus absorbed into modified data-qubit tensors.
Concretely, for $a=1$, each data qubit $i\in\mathrm{supp}(\bar L_Z)$ replaces the prior weight $P(e_i)$ by $(-1)^{\xi_X(e_i)}P(e_i)$, with the modified tensor denoted as $T_Z^{[i]}$, and similarly for $T_X^{[i]}$ with the $Z$-components for $b=1$.
For $a=b=1$, the data qubit at the intersection of the two supports carries the combined sign $(-1)^{\xi_X(e_i)+\xi_Z(e_i)}$ and is denoted as $T_Y^{[i]}$.
These modified tensors are shown as the red, blue, and purple squares in Fig.~\ref{Fig: QEC Sketch}(c), respectively.
For a given syndrome $\mathbf{y}$, the physical indices of $F^{[s]}$ are fixed and the four partition functions can be evaluated by contracting the corresponding tensor networks.

\subsection{PEPO readout error model}
Since most topological codes are defined on 2D lattices, the readout error model should be first generalized to a projected entangled-pair operator (PEPO) to match the geometry of the syndrome graph.
For binary input and output strings $\mathbf{y}$ and $\mathbf{x}$, a correlated classical readout process can be represented as a double-layered PEPO
\begin{align}
    \mathbf{\Lambda}_{\mathbf{x},\mathbf{y}} = \left|\mathrm{Tr}_{\{l_s, r_s, u_s, d_s\}}\prod_s M^{[s]}_{x_s,y_s,l_s,r_s,u_s,d_s}\right|^2,
\end{align}
where the trace is over the virtual indices $\{l_s, r_s, u_s, d_s\}$ connecting neighboring sites, as shown in Fig.~\ref{Fig: QEC Sketch}(d).
This construction guarantees the positivity of the readout error model.
The learning process of the PEPO readout error model can be implemented by minimizing the same KL divergence formula between the predicted distribution and the empirical distribution from the calibration data $\{\mathbf{x}^{(m)}, \mathbf{y}^{(m)}\}_{m=1}^M$ as in Eq.~\eqref{Equ: KL}.
When calculating the loss function and the gradients, the contraction of a 2D tensor network is required, which can be efficiently performed by the boundary MPS method~\cite{Orus2014}.

\subsection{PEPO readout error mitigation for QEC decoding}
Now suppose the measured syndrome is corrupted by readout errors characterized by the error matrix $\mathbf{\Lambda}_{\mathbf{x}, \mathbf{y}}$, and we observe $\mathbf{x}$ instead of $\mathbf{y}$, then the correct ML decoder should evaluate
\begin{align}
    Z_{c_X,c_Z}(\mathbf{x}) = \sum_{\mathbf{e}\in (c_X,c_Z), \mathbf{y}} P(\mathbf{e}) \mathbf{\Lambda}_{\mathbf{x}, \mathbf{y}}\,\delta\left[\boldsymbol\alpha(\mathbf{e}),\mathbf{y}\right] =\sum_{\mathbf{e}\in (c_X,c_Z)} P(\mathbf{e}) \mathbf{\Lambda}_{\mathbf{x}, \boldsymbol\alpha(\mathbf{e})}.
    \label{Equ: QEC Single Round}
\end{align}
In practice, $\mathbf{\Lambda}_{\mathbf{x}, \mathbf{y}}$ is represented by a PEPO learned from calibration data.
To calculate this partition function, one can fuse the PEPO with the decoder PEPS to sum over the likelihood of every hidden syndrome configuration.
Similar to the readout error-free case, the four signed partition functions $Z^{(a,b)}(\mathbf{x})$ can be extracted from four tensor-network contractions, where the only difference is that the local syndrome tensors $F^{[s]}_{y_s}$ are replaced by the fused tensors $G^{[s]}_{x_s}$ that combine the syndrome constraint and the local readout likelihood, as shown in Fig.~\ref{Fig: QEC Sketch}(e).
Specifically, on each syndrome qubit $s$, one contracts the decoder tensor $F^{[s]}_{y_s}$ with the local PEPO tensor $\Lambda^{[s]}_{x_s, y_s}$ as
\begin{align}
    G^{[s]}_{x_s} = \sum_{y_s=0,1} F^{[s]}_{y_s}\, \Lambda^{[s]}_{x_s, y_s},
\end{align}
where $x_s$ is the observed syndrome value on qubit $s$.
This fusion mechanism is the key step of the method.

In real experiments, the syndrome is typically measured repeatedly to mitigate the effect of readout errors by majority voting~\cite{Fowler2012}.
To further improve the decoding performance, one can also directly integrate the PEPO readout model into the multi-round decoding process.
Assume that the underlying data-error configuration $\mathbf{e}$ is fixed and the syndrome is measured independently for $N_r$ rounds, the exact ML objective for measured syndromes $\{\mathbf{x}^{(t)}\}_{t=1}^{N_r}$ then becomes
\begin{align}
    Z_{c_X,c_Z}\left(\{\mathbf{x}^{(t)}\}\right) = \sum_{\mathbf{e}\in (c_X,c_Z)} P(\mathbf{e}) \prod_{t=1}^{N_r} \mathbf{\Lambda}_{\mathbf{x}^{(t)}, \boldsymbol\alpha(\mathbf{e})}.
    \label{Equ: QEC Multi Round}
\end{align}
If each factor in Eq.~\eqref{Equ: QEC Multi Round} is represented by a PEPO, stacking the $N_r$ readout layers on top of the decoder PEPS produces a three-dimensional tensor network.
For a correlated readout error model with $\chi>1$, the exact contraction couples the syndrome qubits across both space and time, so the bond dimension grows exponentially with $N_r$.

To obtain an efficient approximation, we compute the single-site marginal likelihood for every round $t$ and syndrome site $s$ to treat correlations with other sites in an effective mean-field way, i.e.,
\begin{align}
    L_t^{[s]}(y_s) = \sum_{\mathbf{y}_{\backslash s}} \mathbf{\Lambda}_{\mathbf{x}^{(t)}, \mathbf{y}},
    \label{Equ: QEC Marginal}
\end{align}
where all syndrome variables except $y_s$ are summed out by a 2D contraction.
The contributions from different rounds are then accumulated as
\begin{align}
    L_{\rm total}^{[s]}(y_s) = \prod_{t=1}^{N_r} L_t^{[s]}(y_s),
\end{align}
and fused into the decoder tensor as
\begin{align}
    G^{[s]} = \sum_{y_s=0,1} F^{[s]}_{y_s} L_{\rm total}^{[s]}(y_s).
    \label{Equ: QEC 3D Approx}
\end{align}
From a statistical perspective, this approach takes the full sequence $\{\mathbf{x}^{(t)}\}$ as input instead of the majority-voted syndrome, so it can be viewed as a soft-information decoder that exploits the confidence information in the readout data.

\subsection{Numerical simulations}
\begin{figure*}
    \centering
    \includegraphics[width=\linewidth]{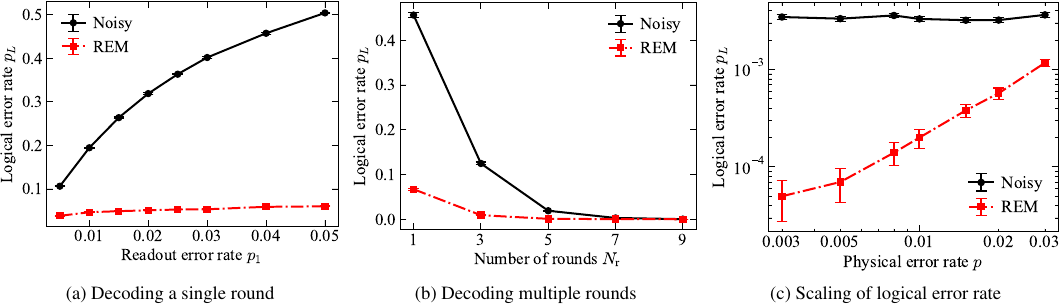}
    \caption{Numerical results for REM in QEC decoding.
    (a) The logical error rate $p_L$ for different readout flip rates $p_1$ (with $p_2=0$) at $p=0.01$ for single-round decoding of $d=7$ with and without REM.
    (b) The logical error rate $p_L$ for repeated-round decoding of $d=7$ with different numbers of rounds at $p_1=0.03$, $p_2=0.005$ and $p=0.01$ with and without REM.
    (c) Scaling of $p_L$ with $p$ for repeated-round decoding of $N_r=d=7$ at $p_1=0.03$ and $p_2=0.005$ with and without REM.}
    \label{Fig: QEC Results}
\end{figure*}

We first consider a simple uncorrelated readout error model with $p_2=0$ acting on a $d=7$ code, which is characterized by a PEPO with $\chi=1$.
In Fig.~\ref{Fig: QEC Results}(a), we compare the logical error rate $p_L$ of the standard decoder without REM and the PEPO-fused decoder with REM for different readout flip rates $p_1$ at a fixed data error rate $p=0.01$.
$p_L$ of the standard decoder increases rapidly as the readout flip rate $p_1$ grows.
By contrast, the PEPO-fused decoder quickly converges to a finite $p_L$, demonstrating the effectiveness of the REM method.

To proceed, we compare the logical error rates for different rounds $N_r$ with and without REM in Fig.~\ref{Fig: QEC Results}(b).
As the number of rounds increases, both decoders benefit from the larger amount of syndrome information, but the PEPO-based decoder improves more significantly because it keeps the full soft-information of the repeated measurements.
By contrast, the decoder without REM effectively compresses each measurement history into a hard syndrome through majority voting and therefore discards the confidence information.
This difference is most visible in the few-round regime, where readout fluctuations are already partially suppressed but still frequent enough that our approach yields a clear advantage.

Furthermore, we study the scaling of the logical error rate $p_L$ with the physical error rate $p$ for multi-round decoding of $N_r=d=7$ in Fig.~\ref{Fig: QEC Results}(c).
If one first compresses the $N_r$ rounds into a hard majority-voted syndrome and then decodes, the resulting logical error rate develops a visible low-$p$ floor where the residual readout errors dominate.
This floor can only be suppressed by increasing the number of rounds $N_r$.
By contrast, the PEPO-marginal decoder continues to suppress the logical error rate as $p$ decreases, largely reducing the required experimental overhead and demonstrating the advantage of the REM method.

\section{Conclusions and Discussions}
We have presented a tensor-network framework for readout error characterization and mitigation based on an MPO representation of the classical readout process.
Compared with conventional uncorrelated models, the MPO ansatz captures local and short-range correlated readout errors while retaining efficient training through tensor contractions.
Using likelihood-based learning from calibration data, we demonstrated that accurate error models can be obtained with a sample cost that grows only near-linearly with system size, and that the resulting readout correlations are non-negligible on real superconducting hardware.

Based on the learned MPO model, we developed REM methods for several important scenarios, including nonlocal observable estimation, global sampling, random-measurement tasks such as classical shadows and learning-based tomography, as well as tensor-network QEC decoding with noisy syndrome readout.
Experiments on superconducting hardware and simulations on large systems show consistent improvements, confirming both practical effectiveness and scalability of the approach.
Our results suggest that tensor-network modeling is a promising direction for systematic treatment of classical noise in quantum measurements.
Moreover, this framework is flexible enough to serve not only as a stand-alone mitigation tool, but also as a building block inside many quantum algorithms and protocols for both near-term and fault-tolerant quantum computing.

One particularly interesting future direction is to compare with the three-dimensional tensor-network decoder used in Google's recent surface-code experiment~\cite{Google2023}, where the spacetime tensor network models circuit-level errors and time-varying syndromes.
Our setting is simpler but complementary, where the true syndrome is fixed while the readout error itself is correlated and learned as a PEPO from calibration data.
A natural next step is therefore to combine these two ingredients and embed a learned PEPO readout layer into a full circuit-level 3D decoder, thereby treating measurement crosstalk and circuit noise within one unified tensor-network inference problem.
However, the resulting 3D tensor network is more challenging to contract, so further algorithmic developments are needed to achieve efficient decoding.
Beyond this specific direction, it would also be valuable to validate the PEPO-fused decoder in real logical-qubit experiments, and to extend the present framework to more general measurement noise beyond the bit-flip channel, such as correlated POVM errors.

\begin{acknowledgments}
    This work is supported by the National Natural Science Foundation of China (NSFC) (Grant No. 12475022 and No. 125B2100) and the Quantum Science and Technology - National Science and Technology Major Project (Grant No. 2021ZD0302100).
\end{acknowledgments}

\bibliography{ref}

\end{document}